\documentclass[11pt]{article}
\usepackage{epsfig}
\usepackage{amssymb}
\usepackage{amsmath}

\newcommand{\ee}{\end{equation}}
\newcommand{\eea}{\end{eqnarray}}
\newcommand{\be}{\begin{equation}}
\newcommand{\bea}{\begin{eqnarray}}

\begin{document}

\title{\LARGE \bf Spherical Structures in Conformal Gravity and its Scalar-Tensor Extension}
  \author{
  \large  Y. Brihaye$^a$ $\:${\em and}$\:$
  Y. Verbin$^b$ \thanks{Electronic addresses: brihaye@umh.ac.be; verbin@oumail.openu.ac.il } }
 \date{ }
   \maketitle
    \centerline{$^a$ \em Physique Th\'eorique et Math\'ematiques, Universit\'e de Mons-Hainaut,}
   \centerline{\em Place du Parc, B-7000  Mons, Belgique}
     \vskip 0.4cm
   \centerline{$^b$ \em Department of Natural Sciences, The Open University
   of Israel,}
   \centerline{\em Raanana 43107, Israel}

\maketitle
\thispagestyle{empty}
\begin{abstract}
We study spherically-symmetric structures in Conformal Gravity and in a scalar-tensor extension and
gain some more insight about these gravitational theories. In both cases we analyze solutions in two systems: 
perfect fluid solutions and boson stars of a self-interacting complex scalar field.  
In the purely tensorial (original) theory we find in a certain domain of parameter space finite mass solutions 
with a linear gravitational potential but without a Newtonian contribution. The scalar-tensor theory exhibits a 
very rich structure of solutions whose main properties are discussed. Among them, solutions with a 
finite radial extension, open solutions with a linear potential and logarithmic modifications and
also a (scalar-tensor) gravitational soliton. This may also be viewed as a {\em static} self-gravitating boson star 
in purely tensorial Conformal Gravity. 

\end{abstract}

\maketitle

\section{Introduction}\label{Introduction}
\setcounter{equation}{0}

Conformal Gravity \cite{Mannheim2006} (CG) was proposed as a possible alternative to Einstein gravity 
(``GR''), which may supply the proper framework for a solution to some of the most annoying problems of 
theoretical physics like those of the cosmological constant, the dark matter and the dark energy. 

It is therefore very much required to investigate its predictions and consequences as further as 
possible. Here we choose to concentrate in localized solutions and to start an investigation of 
their properties. We take two simple matter sources: perfect fluid and complex scalar field, 
we find localized solutions for both kinds of sources and present their main features.

The main ingredient of CG is the replacement of the Einstein-Hilbert action with 
the Weyl action based on 
the Weyl (or {\it conformal}) tensor $C_{\kappa\lambda\mu\nu}$ defined as the totally traceless 
part of the Riemann tensor (we use $R^{\kappa}_{\phantom{\kappa}\lambda\mu\nu}=
\partial_\nu \Gamma^{\kappa}_{\lambda\mu}-\partial_\mu \Gamma^{\kappa}_{\lambda\nu}+...$):
\begin{eqnarray}
C_{\kappa\lambda\mu\nu}=R_{\kappa\lambda\mu\nu}-
\frac{1}{2}(g_{\kappa\mu}R_{\lambda\nu}-g_{\kappa\nu}R_{\lambda\mu}+
g_{\lambda\nu}R_{\kappa\mu}-g_{\lambda\mu}R_{\kappa\nu})+
\frac{R}{6}(g_{\kappa\mu}g_{\lambda\nu}-g_{\kappa\nu}g_{\lambda\mu})
\label{WeylTensor},
\end{eqnarray}
so the gravitational Lagrangian is
\begin{equation}
{\cal L}_{g}= -\frac{1}{2\alpha}C_{\kappa\lambda\mu\nu}C^{\kappa\lambda\mu\nu} 
\label{GravL}
\end{equation}
where $\alpha$ is a dimensionless positive parameter. The gravitational field equations are 
formally similar to Einstein equations where the source is the energy-momentum tensor $T_{\mu\nu}$
and in the left-hand-side Bach tensor $W_{\mu\nu}$ replaces the Einstein tensor:
\begin{equation}
W_{\mu\nu} =  \frac{\alpha}{2} T_{\mu\nu} 
\label{GravFieldEq}
\end{equation}
Bach tensor is defined by:
\begin{eqnarray}
W_{\mu\nu}=\frac{1}{3}\nabla_\mu\nabla_\nu R-\nabla_\lambda\nabla^\lambda R_{\mu\nu}
+\frac{1}{6} (R^2+\nabla_\lambda\nabla^\lambda R-3R_{\kappa\lambda}R^{\kappa\lambda})g_{\mu\nu}+
2R^{\kappa\lambda}R_{\mu\kappa\nu\lambda}-\frac{2}{3}RR_{\mu\nu}
\label{BachTensor}
\end{eqnarray}
Since Bach tensor is traceless, the energy-momentum tensor must ``comply'' so we will consider only
sources with $T^\mu_\mu=0$.

 \vskip 0.5cm
The general spherically-symmetric line-element may be simplified by exploiting the conformal
symmetry and has the form \cite{Mannheim2006}:
\begin{equation}
ds^2= B (r)dt^2 - dr^2/B (r) - r^2 ( d\theta^2+\sin^2 \theta d\varphi^2) 
\label{lineelsph}.
\end{equation}
The non-vanishing components of Ricci tensor and the Ricci scalar are
\begin{eqnarray}
R^0_0 = R^r_r = -\frac{B''}{2}-\frac{B'}{r} \,\,\,\,\,\,; \,\,\,\,\,\, 
R^\theta_\theta = R^\varphi_\varphi = \frac{1-B}{r^2}-\frac{B'}{r} \,\,\,\,\,\,; \,\,\,\,\,\, 
R =\frac{2(1-B)}{r^2} -\frac{4B'}{r}-B''
\label{SphRicci}
\end{eqnarray}
and those of Bach tensor
\begin{eqnarray} \nonumber
W^0_0 = -\frac{1}{3r^4}+ B^2\left[\frac{1}{3r^4}+\frac{1}{3r^2}\left( \frac{B''}{B}+\left( \frac{B'}{B} \right)^2 -
\frac{2}{r}\frac{B'}{B} \right)-\frac{1}{3r}\frac{B'B''}{B^2}+
\frac{1}{12}\left( \frac{B''}{B} \right)^2 \right. \\ \left.
-\frac{1}{6}\frac{B'B'''}{B^2}-\frac{1}{r}\frac{B'''}{B}-\frac{1}{3}\frac{B''''}{B}\right]\\\nonumber
W^r_r =  -\frac{1}{3r^4}+ B^2\left[\frac{1}{3r^4}+\frac{1}{3r^2}\left( \frac{B''}{B}+\left( \frac{B'}{B} \right)^2 -
\frac{2}{r}\frac{B'}{B} \right)-\frac{1}{3r}\frac{B'B''}{B^2}+
\frac{1}{12}\left( \frac{B''}{B} \right)^2 \right. \\ \left.
-\frac{1}{6}\frac{B'B'''}{B^2}+\frac{1}{3r}\frac{B'''}{B}\right]\\\nonumber
W^\theta_\theta = W^\varphi_\varphi = \frac{1}{3r^4}
-B^2\left[\frac{1}{3r^4}+\frac{1}{3r^2}\left( \frac{B''}{B}+
\left( \frac{B'}{B} \right)^2 -\frac{2}{r}\frac{B'}{B} \right)
-\frac{1}{3r}\frac{B'B''}{B^2}+\frac{1}{12}\left( \frac{B''}{B} \right)^2 \right. \\ \left.
-\frac{1}{6}\frac{B'B'''}{B^2}-\frac{1}{3r}\frac{B'''}{B}-\frac{1}{6}\frac{B''''}{B}\right] 
\label{SphBach}
\end{eqnarray}
A useful property of these components is the following:
\begin{equation}
W^0_0-W^r_r=-\frac{B(rB)''''}{3r}\,\,\,\,\,\,; \,\,\,\,\,\, W^r_r+W^\theta_\theta=\frac{B(rB)''''}{6r}.
\label{W00WrrComb}
\end{equation}
In vacuum this is easily integrated to give
\begin{equation}
B(r) = c_0 + c_1 r + c_2 /r + \kappa r^2 \,\,\,\,\,\,; \,\,\,\,\,\, c_0^2=1+3c_1 c_2
\label{ConfSch}
\end{equation}
where the relation between the coefficients comes from the $W^r_r = 0$ equation which is of a third order.
In a non-relativistic fourth 
order gravity a similar situation is encountered, namely the fourth order ``Poisson equation''
\begin{equation}
\nabla^2 \nabla^2 u = -h  
\label{4thOrder}
\end{equation}
where $h(\textbf{r})$ is the source term. In the spherically symmetric case $\nabla^2 \nabla^2 u = (ru)''''/r$
and $u(r)$ is given also by (\ref{ConfSch}) without any relation between the parameters.
On the other hand, the parameters  are related to the source (assumed to extend within $r\leq a$) by
\begin{equation}
c_1 = \frac{1}{2} \int_0^a r^2 h(r) dr \,\,\,\,\,\,; \,\,\,\,\,\, c_2= \frac{1}{6} \int_0^a r^4 h(r) dr .
\label{SourceMoments}
\end{equation}

Since $\kappa$ is not fixed by the source, the $\kappa r^2$ term may be considered as a possible background 
field or in the relativistic context, a cosmological constant contribution\footnote{to be concrete, 
 $R=4\Lambda=-12\kappa$, so $\kappa>0$ corresponds to AdS.}. 
Note also that the volume integral of the matter 
density (i.e. of $h(r)$) turns up as the coefficient of the \textit{linear} term in the potential
rather than the $1/r$ one. It is related to the fact that in this theory the potential of a point particle is
linear in accord with the behavior of the Green function. This linear potential enables one to explain
galaxy rotation curves without assuming dark matter \cite{MannKaz1989,Mannheim2006}. 

For the general case of extended sources, we note that since the field equation is of fourth order, 
a special care should be taken with the boundary conditions.
It is easy to see that $u'(0)$ and $u'''(0)$ should vanish. The value of $u''(0)$ or $u''(\infty)$ may be free 
if solutions with a ``cosmological'' $\kappa r^2$ term are allowed. If on the other hand the background is 
assumed to be empty (``flat''), we may impose further $u''(\infty)=0$ as well. If the source is localized, 
the second derivative at the origin is related to the first moment of the matter distribution as
\begin{equation}
u''(0)=\frac{1}{3} \int_0^a r h(r) dr
\label{u''FirstMom}
\end{equation}  

Now let us return to the relativistic field equations with a perfect fluid source described by 
${T}^{\mu}_{\nu}=diag(\rho,-P_r,-P_\perp,-P_\perp)$  
(with the additional conformal condition $T^{\mu}_{\mu}=0$). Thanks to (\ref{W00WrrComb}) they 
 reduce to a single very simple field equation:
\begin{equation}
\frac{(rB)''''}{r}= -\frac{3\alpha}{2B}(\rho+P_r)
\label{SphGravFieldEqPF}
\end{equation}
which has a similar structure to the fourth order Poisson equation (\ref{4thOrder}). By comparison
we notice that taking $\alpha >0$ corresponds to gravitational attraction in the weak field limit.

Equation (\ref{SphGravFieldEqPF}) should be solved together with the conservation equation
\begin{equation}
P_r'+\frac{1}{r}(3P_r-\rho)+\frac{B'}{2B}(\rho+P_r)=0
\label{ConservEqPF}
\end{equation}
and an additional equation of state which relates algebraically $\rho$, $P_r$ and $P_\perp$. Regularity of the
Bach tensor at the origin introduces an additional boundary condition, $B(0)=1$ to those of the Poisson
case: $B'(0)=B'''(0)=0$, $B''(\infty)=2\kappa$. 

The inertial mass of such a spherical solution is the ordinary 
\begin{equation}
M_I = \int d^{3}x \sqrt{|g|}\,\,\, T^0_0 =  4\pi \int_0^\infty  r^2 \rho (r) dr
\label{IMassPol}.
\end{equation}
However, since the potential of a point particle in this theory is linear, the gravitational mass is identified 
as the coefficient of the linear term in the vacuum potential --
see (\ref{SourceMoments}) and (\ref{SphGravFieldEqPF}):
\begin{equation}
M_G =  12\pi \int_0^\infty dr r^2 (\rho (r)+ P_r (r))/B(r) = \frac{16 \pi}{\alpha} c_1
\label{GMassPol}
\end{equation}
The other parameter, $c_2$ (the coefficient of the $1/r$ term in the potential) has a dimension of length 
which by utilizing Newton's constant can be converted to a mass. However, we don't have an appropriate dimensionful 
parameter at our disposal, so we will call $c_2$  the ``second mass parameter''. In terms of the source functions 
it is given by
\begin{equation}
c_2 =  \frac{\alpha}{4} \int_0^\infty dr r^4 (\rho (r)+ P_r (r))/B(r) \label{NMassPol}
\end{equation}
We will see in the next sections that this integral is not always convergent, and whenever it does, it has the 
wrong sign for an attractive force, causing a non-Newtonian ``near field'' of such sources.  Actually,
 this problem that ordinary continuous sources do not produce a Newtonian component in CG
 was noted already by Mannheim and Kazanas \cite{MannKaz1994} (following even earlier studies 
 \cite{Buchdahll1962,PechSexl1966,DeserEtAl1974,Havas1977}  from the 1960's and 70's).
Mannheim and Kazanas pointed  out towards a possible solution based on the fact that a highly singular source can 
produce a potential with both $c_1>0$ and $c_2<0$. 
 Still, when the implications and consequences of CG are analyzed, smooth matter distributions should
 be considered and studied since they are more widely used to model astrophysical and cosmological sources.

\section{Spherically-Symmetric Perfect Fluid Solutions}
\setcounter{equation}{0}

In accord with our objective, which is investigating the properties of self gravitating solutions in 
CG, we solved Eqs. (\ref{SphGravFieldEqPF}), (\ref{ConservEqPF}) for a set of matter 
distributions.

The simplest of all sources is a constant energy density, $\rho (r)=\rho_0$ (for $r\leq a$ and 0 outside),
but unlike the Einsteinian case, there are no finite mass solutions of this kind in our case. 

The ``next to simplest'' source is a polytrope - either linear with $P_r = \rho/n$ or non-linear (and anisotropic) 
with  $P_r = P_0(\rho/3P_0)^\gamma $ where $n$, $\gamma$ and $P_0$ are all positive constants. The parameter
 $P_0$ is indeed the central value of the pressure (if $P_r(0)$ is finite). 
 Note that the special value $\gamma = 1$ gives only the 
$n = 3$ case of the linear relation which corresponds to isotropic radiation. The other values of $n$ cannot
be obtained as a limit of the non-linear polytrope.

 \vskip 0.5cm
Next we move to general polytropes, that is, density and pressure related by
\begin{equation}
     \rho = 3P_0 (P_r /P_0 )^{1+A}
 \label{RadPolytrope}    
\end{equation}
where for convenience we parametrize the polytropic index by $1/\gamma=1+A$.
The construction of regular solutions for $r \in [0, \infty]$ requires the boundary
conditions 
\begin{equation}
B(0) = 1 \ , \ B'(0) = 0 \ , \ B'''(0) = 0 \ , \ P_r(0) = P_0 \ ,
\end{equation}
the fifth boundary condition was fixed by imposing the value $B''(\infty)$ which is related to the 
free ``cosmological constant parameter'' $\kappa$ (see (\ref{ConfSch})). 
The numerical results further indicate that the solutions behave asymptotically according to
   \begin{equation}
        B = \kappa r^2 + B_1 r + B_0 + \dots \ \ \  , \ \ \ P_r \propto r^{-p}
   \end{equation}
   where the constant $p$ depends on $\kappa$ and on $A$. 

We will discuss separately the solutions available for vanishing and non vanishing $\kappa$, that is 
$B''(\infty)=0$ and $B''(\infty) \neq 0$.

\vskip 2.0cm
\noindent
{\bf Solutions with $\kappa=0$}
\vskip 0.5cm

By examining the conservation equation (\ref{ConservEqPF}), we obtain 
the physically acceptable decay of the function $P_r(r)$ in terms of the 
parameter $A$. It turns out that
solution with an  asymptotically decreasing  $P_r$ can only occur for $A \geq 0$. We then get
$P_r \sim r^{-2}$ for $A=0$ and $P_r \sim r^{-7/2}$ for $A > 0$.
From these observations, it follows that the inertial and gravitating mass given by (\ref{IMassPol}) and 
(\ref{GMassPol}), do not converge in the case $A=0$ (``radiation ball''). Constant density solutions ($A=-1$)
do not exist as well.

\begin{figure}[!b]
\centering
\leavevmode\epsfxsize=10.0cm
 \includegraphics[height=.26\textheight,width=.48\textwidth]{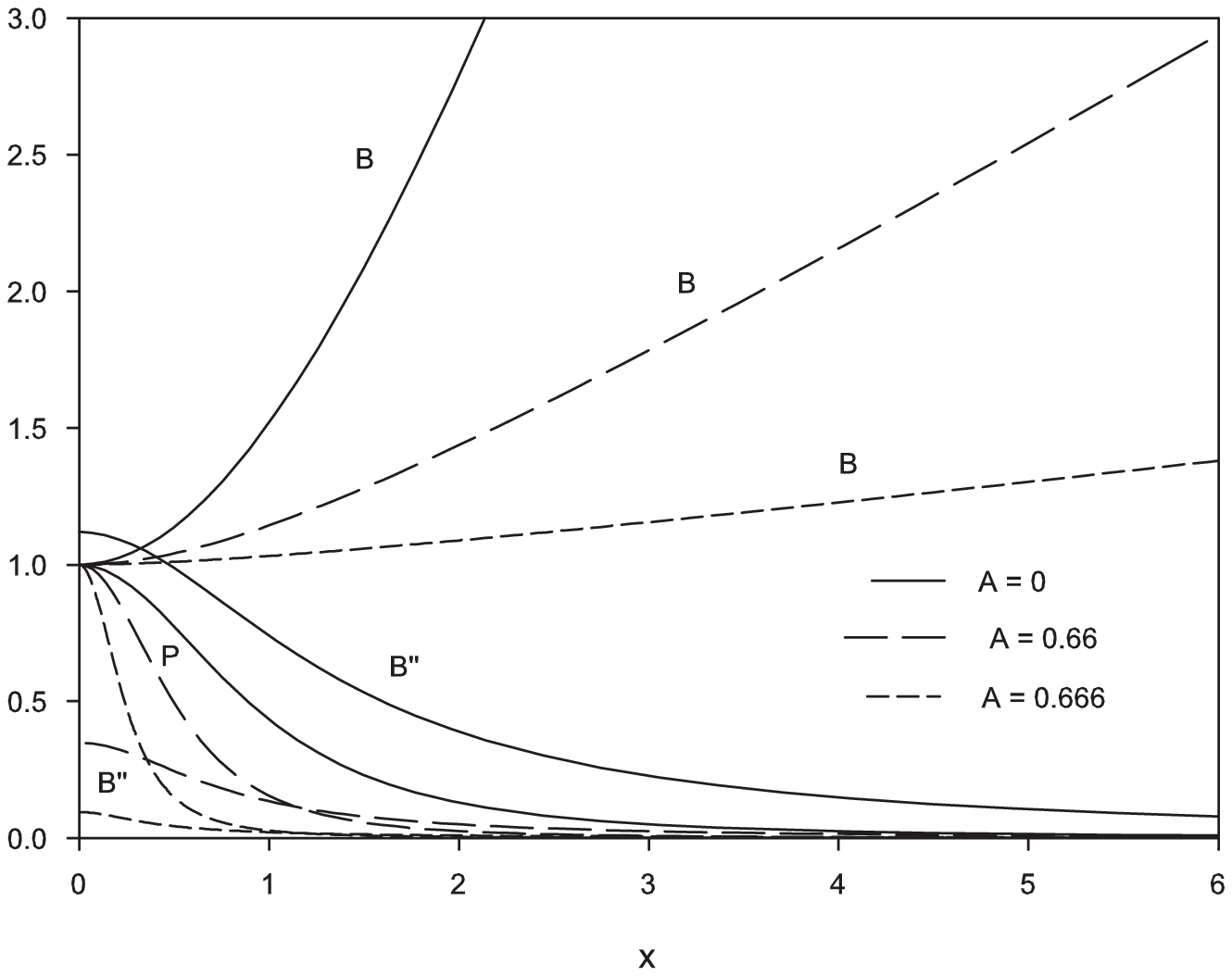}  
 \includegraphics[height=.26\textheight,width=.48\textwidth]{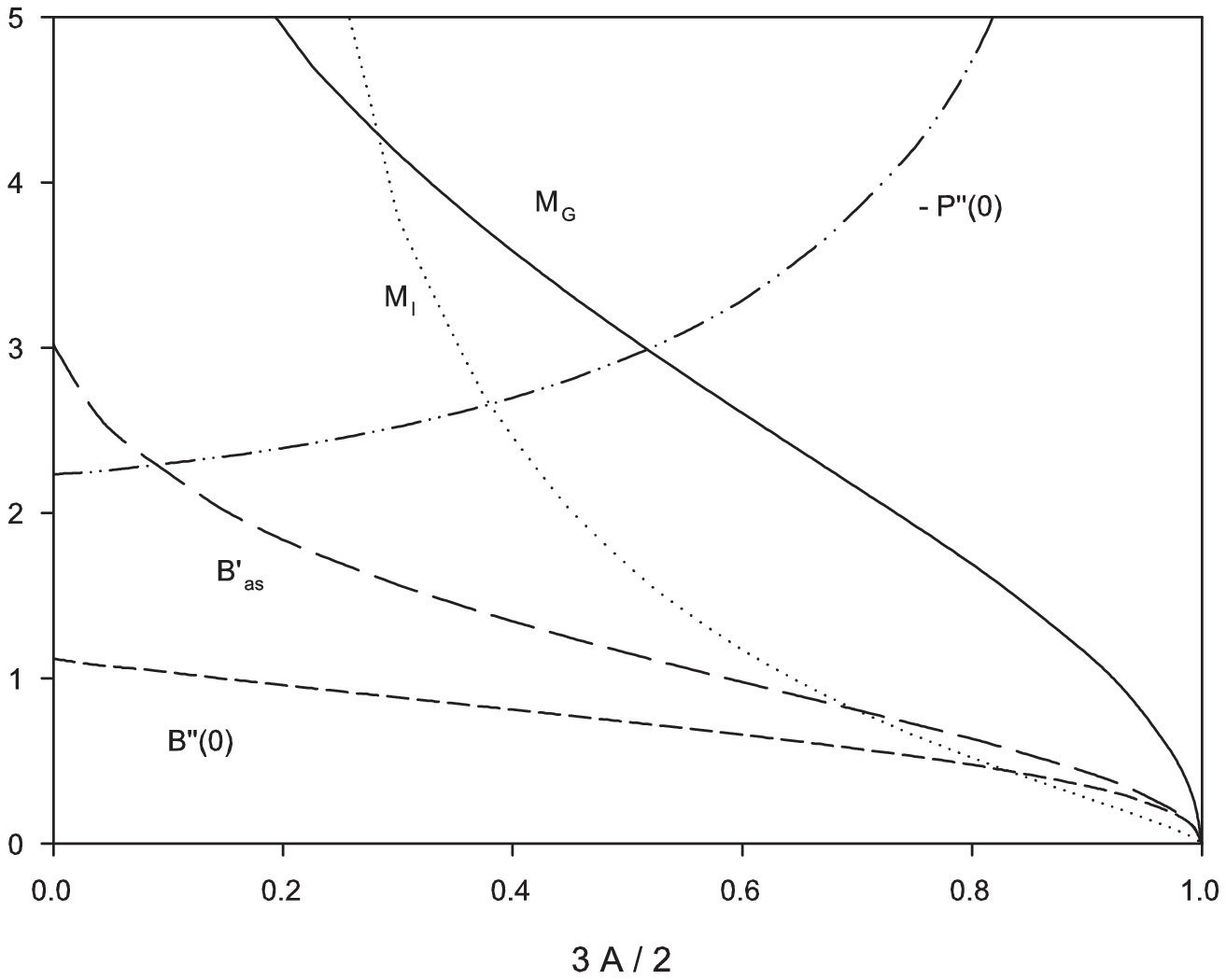}\\
 (a)\hskip 7.5cm (b)\\
\caption{\label{poly12}  \small{ Perfect fluid solutions with $\kappa =0$ :
(a) the profiles of three solutions; 
(b) plots of several characteristics of the solutions as a function of the parameter $A$. }}
\end{figure}


Expanding Eq. (\ref{ConservEqPF}) around the origin, we further observe the following relation
\begin{equation}
      (1 - \frac{3}{2} A)\frac{P_r''(0)}{P_0} + 2 B''(0) = 0 
 \label{constraint}     
\end{equation}
 suggesting that the value $A = 2/3$ should play a role in the solutions.
 Integrating the equations we obtained finite mass solutions 
  for $0 < A < 2/3$. Fig. \ref{poly12}a
 contains  graphic representations of three solutions in this range. Actually, we solved
 a dimensionless version of Eqs. (\ref{SphGravFieldEqPF}), (\ref{ConservEqPF}) and
 (\ref{RadPolytrope}) for 
 $B$, $P_r/P_0$ and $\rho/P_0$ in terms of $x=r(\alpha P_0)^{1/4}$. It is clear from
 the plots that the gravitational potential is asymptotically linear which is the required form in 
 order to explain the galactic rotation curves within this context \cite{MannKaz1989,Mannheim2006}. 
 However, a closer inspection shows that the $1/r$ component, which is necessary for the recovery
 of the Newtonian (Schwarzschild) behavior in smaller scales, is missing. This is reflected by the fact that the  
 coefficient of the $1/r$ term, $c_2$ (see (\ref{NMassPol})) diverges. 
  
 Several parameters characterizing the solutions (namely the masses, the values $B''(0)$,
 $P_r''(0)$ and $B'(\infty)$ are depicted on Fig. \ref{poly12}b; it shows in particular that
 the solution is well defined in the limit $A = 0$. In fact, in this case, we have $P_r = P_0/B^2$ 
 but the masses are infinite. The limit $A \to 2/3$ is more subtle. It seems indeed that in this 
 limit the function $B(r)$ approaches $B=1$ on the full space, while the function $P_r$ becomes more and 
 more concentrated around the origin and  $|P_r''(0)| \to \infty$. We checked that the relation (\ref{constraint})
 is obeyed. At the same time the  masses approach zero.
 
 It is expected to also consider the equations for $A > 2/3$. We were able  to
 obtain solutions in this case. Our numerical results strongly suggest, however
 that no globally regular solutions exist there.  The solutions have $B''(0) < 0$ and the function
 $B(r)$ approaches a zero at some finite value $r=r_0$. At the same time, the pressure
 becomes singular for $r \to r_0$, suggesting that the solution is singular.
 Of course, the numerical construction of such solutions
 cannot be achieved directly for $r \in [0,\infty]$; in fact, we proceed on a small
 interval $r\in [0,r_{max}]$ and gradually increase $r_{max}$.

 \vskip 0.5cm
\noindent 
{\bf Solutions with $\kappa >0$}
 \vskip 0.5cm
The pattern of the solutions is very similar to the case $\kappa =0$.
In particular, regular solutions are also limited to $A < 2/3$ and singularities appear for $A > 2/3$.
It is worth noticing  that the conservation equation (\ref{ConservEqPF}) implies now $P_r \sim r^{-4}$ (for $A \geq 0 $),
so that the masses are finite in the limit $A \to 0$. Of course the values of the masses depend on the 
value adopted for $B''(0)$. Another difference with respect to the case $\kappa =0$, is that now the  
second mass parameter (Eq. (\ref{NMassPol})) converges and a $1/r$ term appears in the gravitational potential, 
but with a wrong sign. Moreover, the convergence is related to the non-vanishing cosmological term $\kappa$, 
so this potential is of a universal nature rather than of local one. 

In the case $\kappa <0$, the field $B(r)$ has a node at a finite $r$ say, $r=r_0$ which leads to a singularity 
of the matter function. This is just the de Sitter horizon which is related to the fact that de Sitter space
does not admit a globally static coordinate system. We will not consider the possibility of asymptotically 
de Sitter space further in this work apart from few mentions.
 \begin{figure}[!t]
\centering
\leavevmode\epsfxsize=10.0cm
 \includegraphics[height=.26\textheight,width=.48\textwidth]{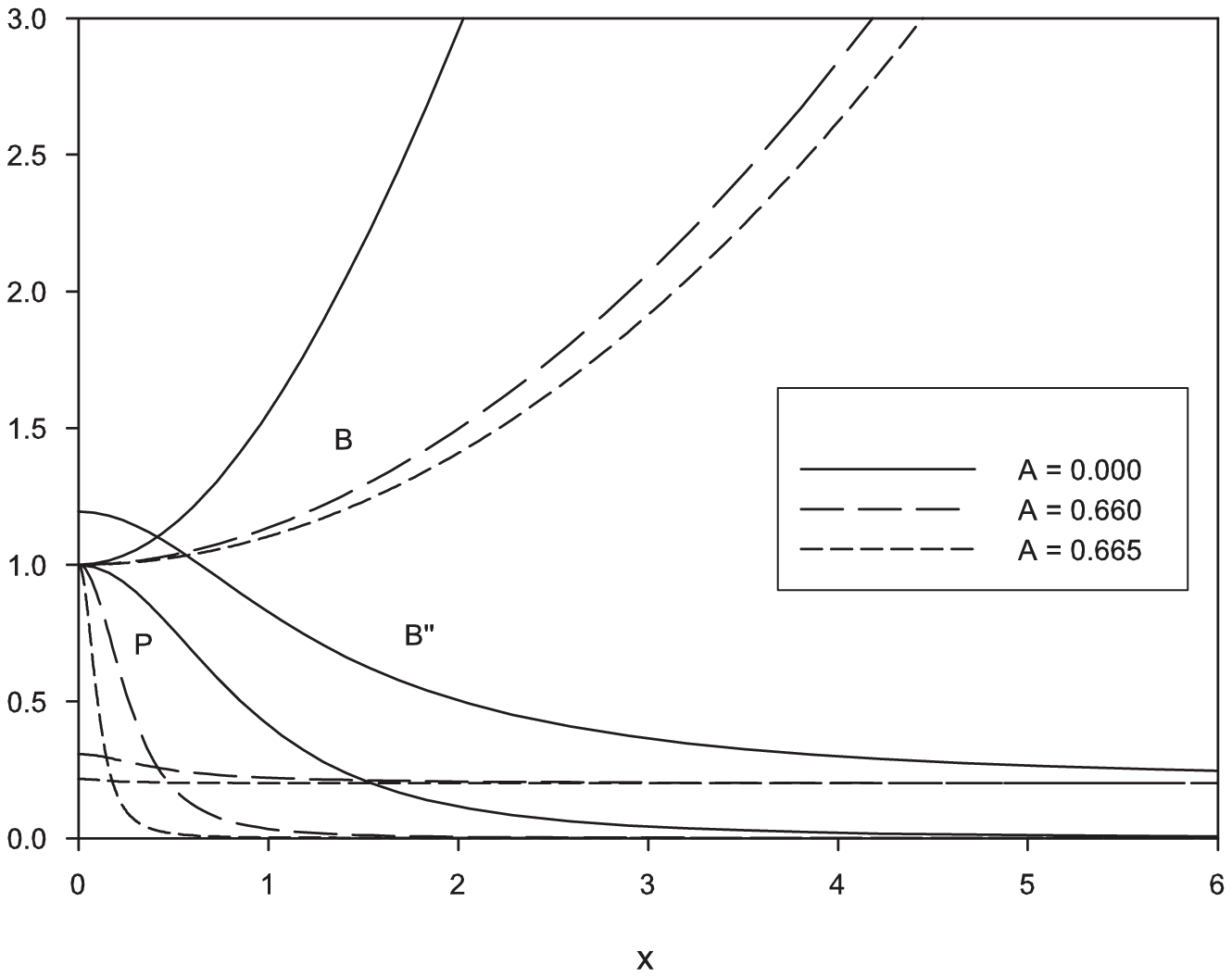}  
 \includegraphics[height=.26\textheight,width=.48\textwidth]{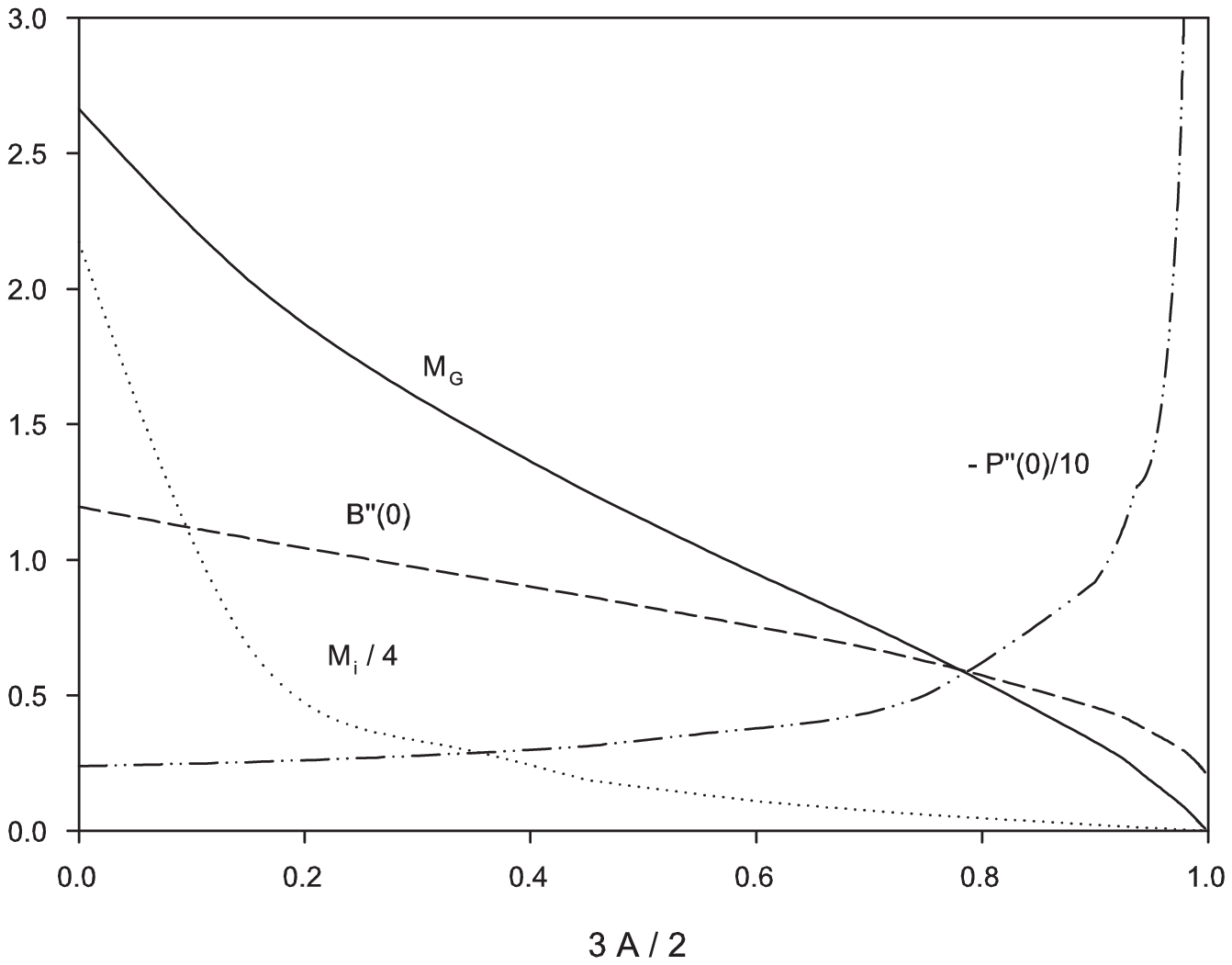}\\
 (a)\hskip 7.5cm (b)\\
\caption{\label{polyKapPos}  \small{Perfect fluid solutions with $\kappa =0.1$:
(a) the profiles of three solutions;  
(b) plots of several characteristics of the solutions as a function of the parameter $A$. Note that unlike
the $\kappa =0$ case, here $B''(0)$ does not vanish as $A\rightarrow 2/3$.}}
\end{figure} 

\section{Boson Stars}\label{BS}
\setcounter{equation}{0}

Among all the higher order gravitational theories \cite{Schmidt2006,Fabbri2008}, CG 
is unique in the sense that
it is based on an additional symmetry principle. The conformal symmetry imposes severe limitations 
on the allowed matter sources. When matter is described in terms of an energy-momentum tensor
it should be traceless as mentioned above already. Similarly the matter Lagrangian is very much
constrained, but the Abelian Higgs model is essentially 
still consistent with the conformal symmetry provided the scalar field ``mass term'' is replaced 
with the appropriate ``conformal coupling'' term which introduces a coupling to the Ricci scalar 
$R$. The matter Lagrangian which we will use here is therefore
\begin{equation}
{\cal L}_{m}= \frac{1}{2}(D_\mu \Phi)^*(D^\mu
\Phi)-\frac{1}{12}R|\Phi|^2 -\frac{\lambda}{4}|\Phi|^4 
-\frac{1}{4}F_{\mu\nu}F^{\mu\nu} \label{matterL},
\end{equation}
and the resulting field equations are
\begin{eqnarray}
D_\mu D^\mu \Phi + \lambda |\Phi|^2 \Phi + \frac{R}{6} \Phi &=& 0 \label{FieldEqsScalar}\\
\nabla_{\mu } F^{\mu\nu} =-\frac{ie}{2}[\Phi^*(D^\nu \Phi)-
\Phi (D^\nu \Phi)^*]&=& J^{\nu }
\label{FieldEqsVector}.
\end{eqnarray}

The gravitational field equations are (\ref{GravFieldEq}) with
\begin{equation}
{T}_{\mu\nu} = { T}_{\mu\nu}^{(minimal)}+{1\over 6}
\left (g_{\mu \nu} \nabla ^{\lambda }\nabla_
{\lambda } |\Phi|^2 - \nabla _{\mu }\nabla_{\nu } |\Phi|^2  - 
{G}_{\mu \nu} |\Phi|^2 \right)
\label{confTmn}
\end{equation}
${T}_{\mu\nu}^{(minimal)}$ being the ordinary (``minimal'') energy-momentum tensor of the Abelian
Higgs model and ${G}_{\mu \nu}$ is the Einstein tensor.

The simplest spherically-symmetric localized solution of this system is the boson star 
\cite{Jetzer1992,LeePang1992,{Liddle1992}} which requires a global U(1) symmetry only - 
that is  $A_{\mu}=0$ and $\Phi=f(r)e^{i\omega t}$. This yields a global conserved charge
which is responsible for its existence.
 
The components of the energy-momentum tensor are (after use of the $\Phi$-equation (\ref{FieldEqsScalar})):
\begin{eqnarray} 
T^0_0 = \frac{5}{6}\frac{\omega^2 f^2}{B}+\frac{B}{6}f'^2-\frac{\lambda}{12}f^4+\frac{B'}{12}(f^2)'+
\frac{1}{18}\left( B''+\frac{B'}{r} + \frac{1-B}{r^2}\right)f^2 \label{SphT00}\\
T^r_r = -\frac{1}{2}\frac{\omega^2 f^2}{B}-\frac{B}{2}f'^2+\frac{\lambda}{4}f^4- 
\frac{1}{12}\left(B'+\frac{4B}{r}\right)(f^2)'-
\frac{1}{6}\left( \frac{B'}{r} - \frac{1-B}{r^2}\right)f^2 \label{SphTrr}\\ 
T^\theta_\theta = T^\varphi_\varphi = -\frac{1}{6}\frac{\omega^2 f^2}{B}+\frac{B}{6}f'^2-
\frac{\lambda}{12}f^4 +\frac{B}{6r}(f^2)'-
\frac{1}{18}\left( \frac{1}{2} B''-\frac{B'}{r} + \frac{2(1-B)}{r^2}\right)f^2 \label{SphTtr}\\
\nonumber
\end{eqnarray}

Since there is only one independent metric component, it is obvious that not all the field equations 
(\ref{GravFieldEq}) are independent. Actually there is only one independent equation and we may use the 
third order one
\begin{equation}
W^r_r - \frac{\alpha}{2} T^r_r = 0
\label{ThrdOrdr}.
\end{equation}
However, a much simpler form is again obtained by using 
(\ref{W00WrrComb}) giving therefore the following fourth order equation for the metric component $B(r)$:
\begin{equation}
\frac{(rB)''''}{r}= -\frac{\alpha}{B}\left[ \frac{2\omega^2 f^2}{B}+ Bf'^2 -\frac{\lambda}{2}f^4 
+\frac{1}{4}\left(B'+\frac{2B}{r}\right)(f^2)'-\frac{R}{12}f^2
\right]
\label{BSFieldEq}.
\end{equation}
For the scalar field we have the second order equation 
\begin{equation}
\frac{\left(r^2Bf'\right)'}{r^2}+
\left(\frac{\omega^2}{B} - \frac{R}{6}\right)f -\lambda f^3 =0 
\label{SphScFieldEq}
\end{equation}
where one should also write explicitly $R =2(1-B)/r^2 -4B'/r-B''$ by (\ref{SphRicci}).

The inertial mass and gravitational mass of these boson stars are given by equations
like (\ref{IMassPol}) and (\ref{GMassPol}) with the necessary adaptations:
\begin{equation}
M_I =  4\pi \int_0^\infty  dr r^2 T^0_0 (r) 
\label{IMassBS}
\end{equation}
\begin{equation}
M_G =  12\pi \int_0^\infty dr r^2 (T^0_0 (r)- T^r_r (r))/B(r) 
\label{GMassBS}
\end{equation}
where $T^0_0 $ and $T^r_r $ are given by (\ref{SphT00}) and (\ref{SphTrr}). The
second mass parameter is defined in analogy with (\ref{NMassPol}):
\begin{equation}
c_2 =  \frac{\alpha}{4} \int_0^\infty dr r^4 (T^0_0 (r)- T^r_r (r))/B(r)
\label{NMassBS}
\end{equation}
The boson star has also a global charge (particle number) which is given by
\begin{equation}
Q =  4\pi \omega \int_0^\infty dr r^2 f^2(r)/B(r) 
\label{ChargeBS}.
\end{equation}

It is interesting to note that the field equations form an autonomous system as a result of
 the transformation
\begin{equation}
B(r)=V (r)r^2 \,\,\,\,\,\,; \,\,\,\,\,\, f(r)=\varphi (r) /r \,\,\,\,\,\,; \,\,\,\,\,\,
r=1/u
\label{AutonTransf}
\end{equation}
and they also simplify considerably:
\begin{eqnarray}
\frac{1}{\alpha}V'''' + (\varphi')^2 - \frac{1}{2}\varphi\varphi'' + \frac{3 \omega^2 \varphi^2}{2V^2}= 0 \\
\left(V\varphi'\right)'+ \frac{\omega^2}{V} \varphi +
 \frac{V''-2}{6}\varphi -\lambda \varphi^3 = 0 .\nonumber 
\label{AuronSphScFieldEq}
\end{eqnarray}
Here of course $'=d/du$. These field equations may be obtained from the following ``reduced Lagrangian'':
\begin{equation}
L_{red} = \frac{1}{6\alpha}(V'')^2+
\frac{V}{2}  (\varphi ')^2-\frac{\omega ^2 \varphi ^2}{2 V}
+\frac{1}{6} \left(1-\frac{V''}{2}\right)\varphi ^2
  +\frac{\lambda}{4}  \varphi^4
\label{RedLag}
\end{equation}
There is also a ``conserved energy'' $K$ (such that $K' = 0$) :
\begin{equation}
K =\frac{1}{6\alpha}\left( (V'')^2-2 V' V''' \right) +
\frac{V}{2}  (\varphi ')^2+ \frac{\omega ^2 \varphi ^2}{2 V}
 +\frac{V'}{6}\varphi\varphi' -\frac{1}{6} \varphi ^2 
  -\frac{\lambda}{4}  \varphi^4
\label{HamiltonianAut}
\end{equation}
whose value is not free but fixed to be $K=2/3\alpha$ since Eq. (\ref{HamiltonianAut})
is equivalent to (\ref{ThrdOrdr}). Moreover, if we define a third degree of freedom $W$, such that $W=V''$,
the equations of motion can be derived from the
following ``ordinary'' second order Lagrangian
\begin{equation}
L_{2} = \frac{V}{2}  (\varphi ')^2-\frac{\omega ^2 \varphi ^2}{2 V}
+\frac{1}{6} \varphi ^2   +\frac{\lambda}{4}  \varphi^4 +\frac{1}{6} V' \varphi \varphi' 
-\frac{1}{6\alpha}(W^2+2V'W')
\label{Lag2ndOrder}
\end{equation}

\section{Boson Stars: Numerical Results}\label{BSNum}
\setcounter{equation}{0}

In absence of explicit solutions (not even to the simple autonomous system), we approached the system of 
equations (\ref{BSFieldEq}), (\ref{SphScFieldEq})
numerically. Using an appropriate rescaling  $r\to C r$ and  $f\to F f$, the coupling constants
$\lambda$, $\alpha$ scale with a factor $C^2 F^2$ while $\omega$ scales by $C$. 
Using these rescaling, we can  set $\omega = \alpha =1$
in the equations and study the solutions for several values of the coupling constant $\lambda$.
If we denote by $\tilde{f}(x)$ and $\tilde{B}(x)$ the solution with $\omega = \alpha =1$ and a given $\lambda$, the
solutions with  general values of $\omega$ and $\alpha$ and self-coupling $\alpha\lambda$ are
\begin{equation}
f(r) = \frac{\omega}{\sqrt{\alpha}} \tilde{f}(\omega r) \,\,\,\,\,\,; 
\,\,\,\,\,\, B(r) =  \tilde{B}(\omega r) .
\label{GenProfiles}
\end{equation} 
It is also easy to see that the charge $Q$ is independent of the parameter $\omega$ and the mass scales like 
$\omega/\alpha$.

Since we chose to solve the fourth order equation (\ref{BSFieldEq}), Eq. (\ref{ThrdOrdr}) serves as 
a constraint. Taking the derivatives of the left-hand side of (\ref{ThrdOrdr}) with respect to $r$ 
and eliminating the maximal derivatives $B''''$ and $f''$ by using (\ref{BSFieldEq})-(\ref{SphScFieldEq}),
leads to an expression which vanishes identically. This implies that the two equations we chose to solve guarantee
that the combination $W^r_r - \frac{\alpha}{2} T^r_r $ is constant so the constraint will be automatically 
fulfilled (for any $r$) by a consistent choice of the boundary conditions (such that the constant is 0). 

We first discuss the solutions in the case where the function $B(r)$ is asymptotically linear;
 that is to say $B''(\infty) = 0$, or $\kappa = 0$ in (\ref{ConfSch}). The relevant set of boundary conditions for 
 solutions of this type is
\begin{equation}
B(0)=1 \ \ , \ \ B'(0)=0 \ \ , \ \ B'''(0) = 0 \ \ , \ \ f'(0) = 0 \ \ , \ \ B''(\infty) = 0 \ \ , \ \ f(\infty)=0
\end{equation}
For a better understanding of the numerical results,
it is  instructive to analyze the asymptotic possible behavior of the solutions. The asymptotic form of the $B$
field, i.e. $B(r) \sim B_1 r$, enforces the function $f(r)$ to obey asymptotically an hypergeometric equation
whose solutions are of the form
\begin{equation}
  f(r) = \frac{F_0}{r} \sin \left( \frac{\log(\omega r)}{B_1} + \varphi \right) \ \ , \ \ r \to \infty
\end{equation}
where $F_0, \varphi$ are constants.
As a consequence, the function $f(r)$ oscillates asymptotically and necessarily develops nodes,
 rendering  the numerical integration  technically difficult. We manage however to construct the 
 solution by replacing the condition $f(\infty)=0$ by $f(r_0)=0$ imposing  by hand  the
 first zero $r_0$ of the function $f(r)$.
 Proceeding this way, we obtained strong numerical evidences that a continuum family of solutions
 exist, labelled by $r_0$. In particular, the values $B''(0)$, $f(0)$, $B_1$ are fixed by $r_0$. Unfortunately,
 the  integrated energy density and particle number densities of these solutions behave according to
 \begin{equation}
  \int dr  \frac{1}{r} \left(\sin\left({\log(\omega r) /B_1 + 
  \varphi}\right)\right)^2 \sim \int dy \sin^2( y + \varphi)  
  \ \ ,   y = \log(\omega r) \ \ .
\end{equation}
The corresponding mass and particle number are then infinite. In other words, there do not exist in this theory
boson star solutions with a linear gravitational potential.

On the other hand, boson stars exist with a quadratic gravitational potential which corresponds to an asymptotically
anti de Sitter space (a negative cosmological constant). 
Setting $B''(\infty) = 2 \kappa$, we obtain the
   asymptotic form
   \begin{equation}
             B(r) = \kappa r^2 + B_1 r + B_0 + \dots \ \ \ , \ \ \ f = f_1/r + f_2/r^2 + \dots
   \end{equation}
   
\begin{figure}[!t]
\centering
\leavevmode\epsfxsize=10.0cm
 \includegraphics[height=.26\textheight,width=.48\textwidth]{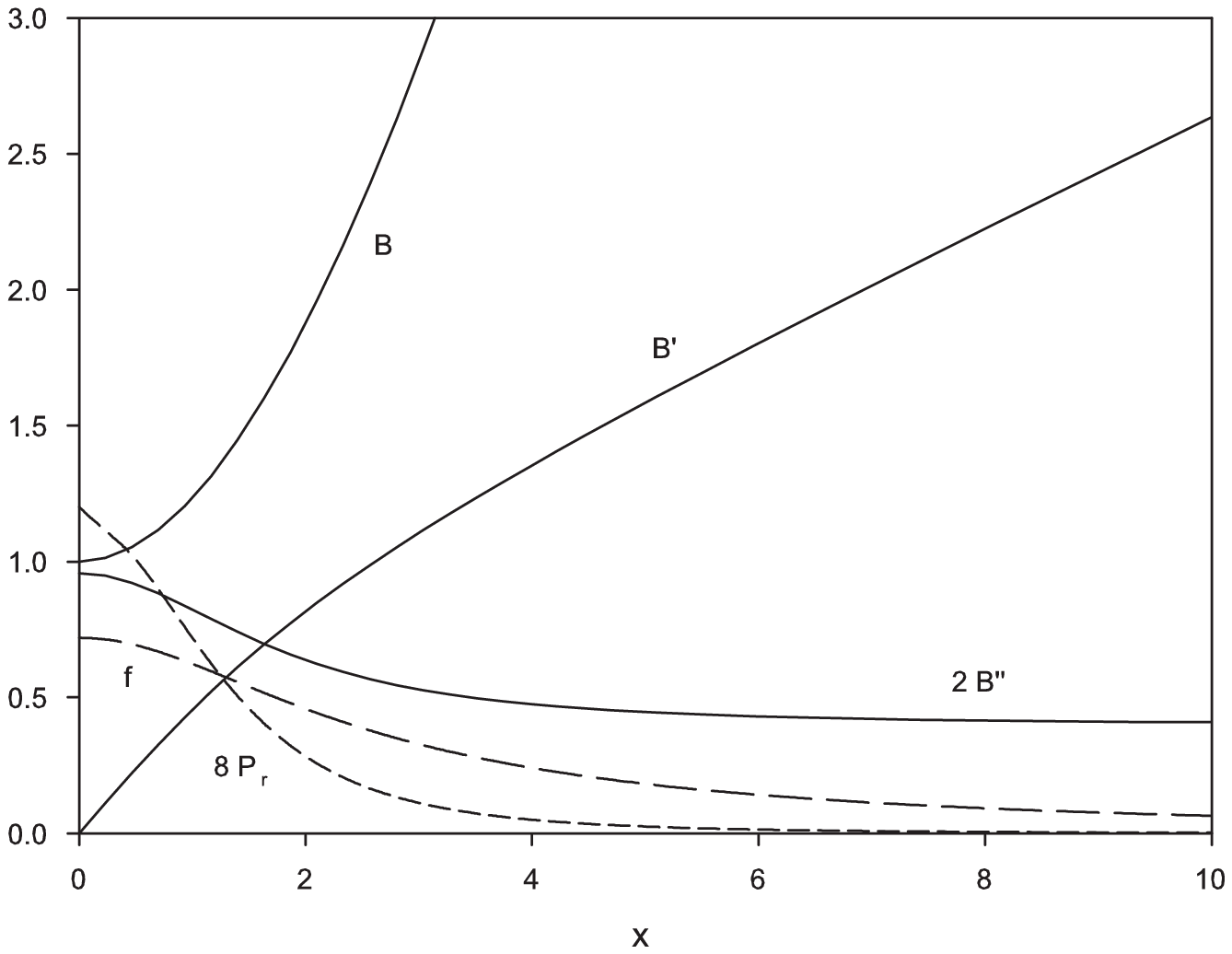}  
 \includegraphics[height=.26\textheight,width=.48\textwidth]{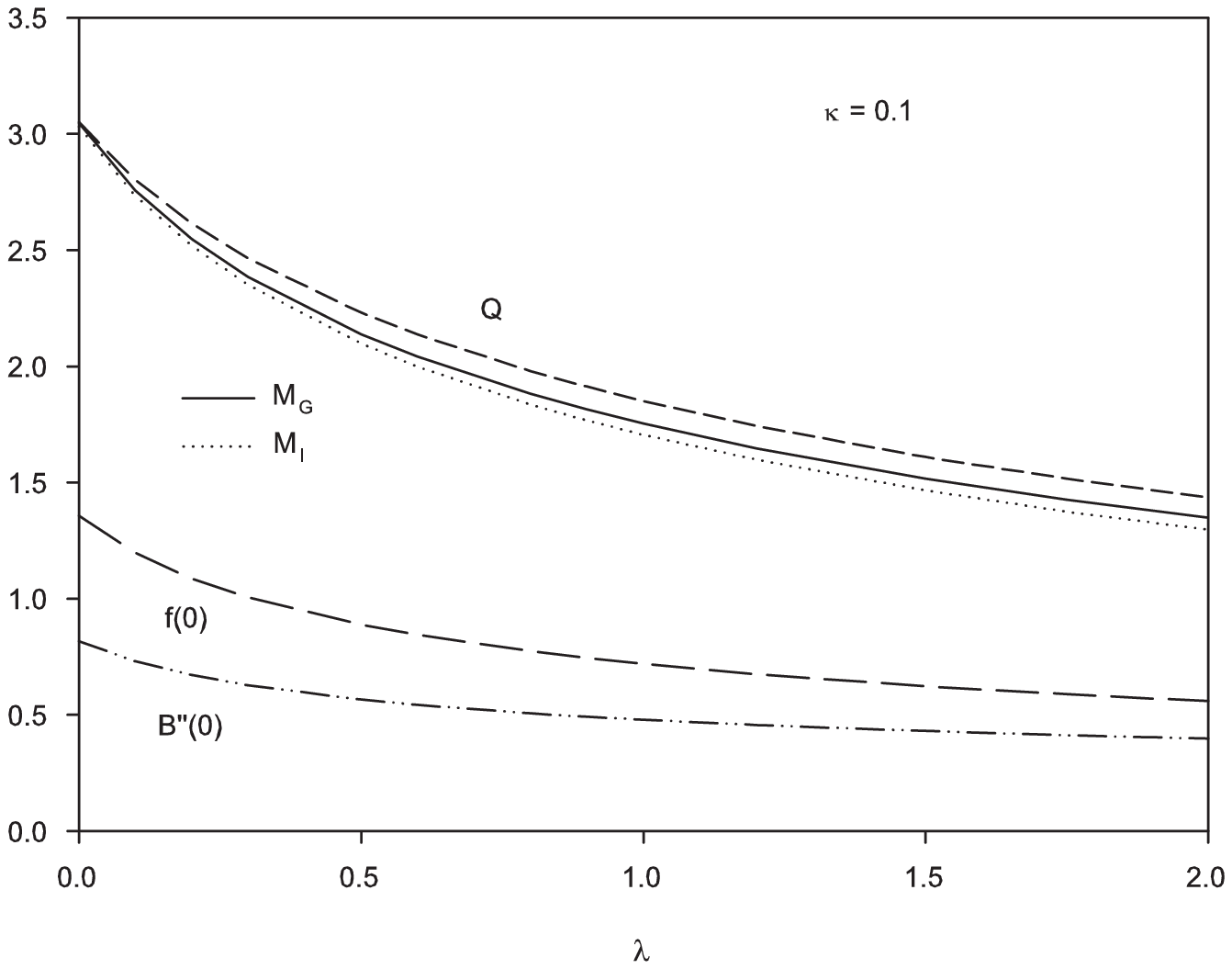}\\
 (a)\hskip 7.5cm (b)\\
\caption{\label{cos_cons-phys_dat}  \small{Boson stars with $\kappa=0.1$.
(a) the profiles of a typical conformal boson star solution with $\lambda=1$; 
(b) plots of several characteristics of the solutions as a function of $\lambda$:
The inertial mass $M_I$ and particle number $Q$ are plotted in units $8\pi$ while the gravitational mass 
$M_G$ is given in units $4 \pi$. }}
\end{figure}    
   

Finite mass solution needs to impose the stronger decay at infinity such that $f_1=0$.
We obtained strong numerical evidences that such solutions exist, that is we solved numerically the
field equations for a wide range of the self-coupling parameter $\lambda$. A typical profile is presented
in Fig. \ref{cos_cons-phys_dat}a with $\lambda=1$ and  $\kappa = 0.1$. 
Several physical characteristics of the solutions are plotted for $\lambda \in [0,2]$
in Fig. \ref{cos_cons-phys_dat}b.  Surprisingly, the solutions seems to persist in the absence of 
self-interaction ($\lambda = 0$).

Due to the asymptotic behavior mentioned above, the second mass parameter (\ref{NMassBS}) is also finite, namely the
gravitational potential contains in this case too a wrong sign $1/r$ term.

\section{Scalar-Tensor Conformal Gravity}
\setcounter{equation}{0}

CG has been criticized from several aspects both phenomenological and formal. 
Several authors claim that predictions in the weak field limit disagree 
with solar system observations \cite{Flanagan}, yield wrong light deflection \cite{EderyPar1998} (see however 
suggestions  \cite{Mannheim2007,EderyEtAl2001} for circumventing the difficulties) or
more generally, the exterior solution (\ref{ConfSch}) with $\kappa=0$, $c_1 >0$ and $c_2 <0$ 
(which yields the desired behavior) cannot be matched to any source with a ``reasonable'' 
mass distribution \cite{PerlickXu}. In this respect
we have found in the previous sections that boson stars indeed cannot produce such a behavior. On the other 
hand, the ``anisotropic'' polytropes (Eq. (\ref{RadPolytrope})) may present a linear potential, but the 
$1/r$ component is missing.

Other authors find evidence for tachyons or ghosts 
\cite{BarabashSht1999} or raise the fact that only  null geodesics are physically meaningful in this 
theory since the ``standard'' point particle Lagrangian is not conformally-invariant \cite{WoodMoreau}.

This last point (and possibly some of the former) can be easily corrected and can serve as a starting point for 
a consistent conformal theory by adding a real
scalar field and turning the theory into a scalar-tensor theory. The conformally-invariant point 
particle Lagrangian will be 
\begin{equation}
L_{pp}=-S\sqrt{g_{\mu\nu}\dot{x}^\mu \dot{x}^\nu} 
\label{L-PointP}
\end{equation}
where $S$ is a real scalar field with the usual conformal transformation laws.
The gravitational Lagrangian (\ref{GravL}) will be modified to
\begin{equation}
{\cal L}_{g}= \frac{1}{\alpha}\left(-\frac{1}{2}C_{\kappa\lambda\mu\nu}C^{\kappa\lambda\mu\nu}
+\frac{1}{2}\nabla _{\lambda }S\nabla^{\lambda } S-\frac{1}{12}RS^2 -\frac{\nu}{4}S^4  \right)  
\label{STGravL}
\end{equation}
where $\nu$ is a possible self-coupling parameter.

The field equations will be modified accordingly. First of all, there will be an additional scalar field equation: 
\begin{eqnarray}
\nabla_\mu \nabla^\mu S + \nu S^3 + \frac{R}{6} S = 0 
\label{ScalarGEq}.
\end{eqnarray}
Second, we turn to the tensorial equations (\ref{GravFieldEq}). Technically, the modification is just an additional 
energy--momentum tensor ${S}_{\mu\nu}$ in the right hand side of (\ref{GravFieldEq}) namely
\begin{equation}
W_{\mu\nu} =  \frac{\alpha}{2} T_{\mu\nu} + \frac{1}{2}{S}_{\mu\nu}
\label{STGravFieldEq}
\end{equation}
where
\begin{equation}
{S}_{\mu\nu} = \partial_\mu S \partial_\nu S - 
g_{\mu\nu}\left ( \frac{1}{2}\nabla _{\lambda }S\nabla^{\lambda } S-\frac{\nu}{4}S^4 \right) 
+{1\over 6}\left (g_{\mu \nu} \nabla ^{\lambda }\nabla_
{\lambda } S^2 - \nabla _{\mu }\nabla_{\nu } S^2  - 
{G}_{\mu \nu} S^2 \right)
\label{Smn}
\end{equation}
But in principle the scalar field should be considered as a gravitational degree of freedom which is stressed by the
absence of the coupling constant $\alpha$ in front of ${S}_{\mu\nu}$.

The simplest case to be studied is static spherically-symmetric vacuum solutions which within this framework are
obtained by solving the following simplified version of Eqs. (\ref{BSFieldEq})-(\ref{SphScFieldEq}) for 
$B(r)$ and $S(r)$:
\begin{eqnarray}
\frac{(rB)''''}{r}+\frac{1}{B}\left[ BS'^2 -\frac{\nu}{2}S^4 
+\frac{1}{4}\left(B'+\frac{2B}{r}\right)(S^2)'-\frac{R}{12}S^2
\right]= 0 \nonumber \\
\frac{\left(r^2BS'\right)'}{r^2} - \frac{R}{6} S -\nu S^3 = 0 
\label{STVacFieldEq}
\end{eqnarray}
where as before, $R$ should be expressed in terms of $B(r)$ using (\ref{SphRicci}). Since the equation
\be 
W^r_r = (1 /2) S^r_r 
\label{ST3rdOrder}
\ee
is of third order, it will be necessary to assure its validity
and this will be done as before by using consistent boundary conditions.

The difference with respect to the boson stars discussed above, is that now we may allow singular solutions
in analogy with the Schwarzschild solution of standard GR. The no-hair theorem which precludes black holes 
with scalar hair is evidently not applicable in the present context.

Actually, one may prefer to study the system in a different gauge where by conformal transformation
the scalar field is a constant, $S(x^\mu)=S_0$. This simplifies considerably the general field equations
(\ref{ScalarGEq})-(\ref{Smn}) and gives immediately the result $R=-6\nu S^2_0$. However, after transforming 
to a constant $S(x^\mu)$, one cannot use the ``Mannheim gauge'' (Eq. (\ref{lineelsph})) any more. 
The metric tensor will have two independent
components and the relatively simple expressions for Bach tensor $W^\mu_\nu$ will become quite cumbersome.

We therefore chose to stick to the ``Mannheim gauge'' and to use $S$ as a second degree of freedom. On the
other hand the ``effective metric'' that a point particle experiences is $\bar{g}_{\mu\nu}=S^2g_{\mu\nu}$ -- see
(\ref{L-PointP}). Consequently, the interpretation of the solutions is now quite different: it is now 
$\bar{g}_{\mu\nu}$ which has the physical significance, and the question of the gravitational
potential should be answered by analyzing $\bar{g}_{00}=S^2(r)B(r)$ rather than $B(r)$. 

As for the purely tensorial case with a scalar field, the vacuum scalar-tensor theory yields an
autonomous system as well. We repeat the transformation (\ref{AutonTransf}) now with $\Sigma(u)=S(1/u)/u$ 
and get the equations of motion
\begin{eqnarray}
V'''' + (\Sigma')^2 - \frac{1}{2}\Sigma\Sigma'' = 0 \nonumber \\ 
\left(V\Sigma'\right)'+  \frac{V''-2}{6}\Sigma -\nu \Sigma^3 = 0  
\label{AutonS-T}
\end{eqnarray}
the ``reduced Lagrangian''
\begin{equation}
L_{ST} = \frac{1}{6}(V'')^2 + \frac{V}{2}  (\Sigma ')^2
+\frac{1}{6} \left(1-\frac{V''}{2}\right)\Sigma ^2
  +\frac{\nu}{4}  \Sigma^4
\label{LagST}
\end{equation}
and the ``conserved energy'' $K_{ST}$ (now $K_{ST}=2/3$):
\begin{equation}
K_{ST} =\frac{1}{6}\left( (V'')^2-2 V' V''' \right)   +\frac{V}{2}  (\Sigma ')^2
+\frac{V'}{6}\Sigma\Sigma'  -\frac{1}{6} \Sigma ^2 -\frac{\nu}{4}  \Sigma^4.
\label{HamiltonianST}
\end{equation}
The second order Lagrangian is in this case just (\ref{Lag2ndOrder}) without the $\omega$ term:
\begin{equation}
L_{2ST} = \frac{V}{2}  (\Sigma ')^2
+\frac{1}{6} \Sigma ^2  +\frac{\nu}{4}  \Sigma^4 +\frac{1}{6} V' \Sigma \Sigma' 
-\frac{1}{6}(W^2+2V'W')
\label{Lag2ndOrderST} 
\end{equation}

\section{Vacuum Solutions}
\setcounter{equation}{0}
\setcounter{table}{0}

\subsection{Schwarzschild-Like Solutions}
Equations (\ref{STVacFieldEq}) possess a three-parameter family of explicit solutions given by
\be
\label{analytic}
  B(r) = ( 1 + r/a)^2 - \frac{r_h}{r} \frac{(1+ r/a)^3}{(1+r_h/a)} 
  + \frac{\nu S_0^2 r_h^2}{2} \Big(\frac{r^2}{r_h^2} - \frac{r_h}{r}\frac{(1+r/a)^3}{(1+r_h/a)^3} \Big)
   \ \ , \ \ S(r) = \frac{S_0}{1 + r/a}
\ee 
where $r_h$, $a$, $S_0$ are free parameters; $a$ and $S_0$
fix respectively the scale of the radial coordinate $r$ and of the scalar function $S(r)$. 
This solution describes a family of black hole 
space-times with a regular horizon at $r=r_h$ satisfying $B(r_h)=0$.

Assuming for simplicity $\nu = 0$, these solutions have the following
behavior:
\be
B'(r_h) = \frac{1}{a}+\frac{1}{r_h} \ \ , \ \   
B(r\to 0) = -\frac{r_h}{(a+ r_h)r}   \ \ , \ \  
B(r\to \infty) =  \frac{r^2}{a(a+r_h)}  + O(r)
\label{BehaviorExplBHSol}   
\ee 
The metric $g_{\mu\nu}$ is therefore asymptotically  anti-de Sitter (or de Sitter) space with
a cosmological constant $\Lambda = - 3/({a(a+r_h)})$.
These formulas  can be generalized for $\nu \neq 0$ but they become more involved;
 the case $\nu = 0$ is sufficient to illustrate our results. 
Fixing the parameters $r_h$ and $a$ (or $\Lambda$), the solutions (\ref{analytic}) can be of three different forms
according to the value of $a$~:
\begin{itemize}
\item $a > 0$~: a single horizon at $r = r_h$.
\item $-r_h < a < 0$~: a regular horizon at $r=r_h$ hidden by a doubly degenerate horizon at
        $r = \tilde r_h > r_h$ with $\tilde r_h = r_h + \sqrt{r_h^2 + 16 \kappa a^4}$. $\Lambda < 0$ in this case.
\item $a < -r_h$~: a regular horizon at $r=r_h$ and, inside, a doubly degenerate horizon at
        $r = \tilde r_h < r_h$ with $\tilde r_h = r_h - \sqrt{r_h^2 + 16 \kappa a^4}$. $\Lambda > 0$ in this case.
\end{itemize}



Note however that although $B(r)$ is quite similar to the solutions of the purely tensorial CG,
the actual gravitational field ``felt'' by a point particle is very different, since the components of the relevant 
metric tensor are those of $\bar{g}_{\mu\nu}=S^2g_{\mu\nu}$ namely $S^2(r)B(r)$, $S^2(r)/B(r)$ and $r^2S^2(r)$. 
Already here we can notice that $S^2(r)B(r)$ increases with $r$ much less steeply, and actually goes asymptotically 
to a constant. Moreover, the circumferential radius $rS(r)$ is also bounded. These solutions are therefore closed.
On the other hand the limit $a \to \infty$ gives rise to yet another kind of solution with constant $S(r)$ and a 
``purely Schwarzschild'' $B(r)$. It is just a special case of a whole family of open solutions that will be discussed 
below in sec. \ref{nuneq0}.

\subsection{General Black Hole Solutions}

The family of solutions discussed in the previous section are entirely determined by the scale of the scalar 
field and by the value of the horizon $r_h$
and the parameter $a$. In particular, the values of the horizon, of the derivative $B'(r_h)$ and of the cosmological 
constant are  not independent. However, since the equation determining the metric field is of the fourth-order, 
 more general solutions are expected. 
In absence of a  generalization of the explicit form  (\ref{analytic}), we investigated the equations by 
 numerical methods. The first step in this direction consists of establishing the most general set of appropriate
boundary conditions. Prior to this step, the following scale invariance of Eqs. (\ref{STVacFieldEq})
has to be fixed~:
\be
\label{rescaling}
r\to C r \ \ , \ \ S \to  \frac{S}{C} \ \ , \ \ B \to B
\ee
where $C$ is a constant. We will fix this arbitrary scale by imposing a particular
value for $S_h \equiv S(r_h)$. So we define a dimensionless scalar field\footnote{We will still use $S$ for it too.} 
$S/S_h$ and a radial variable $x=r/|a|$. For the vacuum solution (\ref{analytic}), this scale fixing yields
the relation $S_0=S_h(1+x_h)$. 

Solutions presenting a regular horizon at $x = x_h$ require the following conditions~:
\be
\label{bc_scalar}
     B(x_h)=0 \ \ ,  \ B'(x_h) = b \ ,  \ 
     {\cal G}|_{x=x_h} = 0 \ \ ,  \ {\cal H}|_{x=x_h} = 0 \ \ ,  \ S(x_h)=1
     \ \ , \ \ B''(\infty) = B_2 \equiv 2\kappa a^2
\ee
where the symbols $\cal{G,H}$ represent (respectively) the conditions of regularity of  Eq.(\ref{STVacFieldEq}) 
 at the horizon and the constraint (\ref{ST3rdOrder}):
\be
    {\cal G} = 6 B' S' - S(\frac{2}{x_h^2} - \frac{4 B'}{x_h} - B'') - 6 \nu B^3
\ee
\be
     {\cal H} = 2 x_h^2 (4 (B')^2 - S^2) + 2 x_h^3( B' S^2 - 4 B' B'') 
     + x_h^4 ( 2(B'')^2 - 4 B' B'''  + 2 B' S S'- 3 \nu S^4) - 8
\ee
The normalization chosen for the field $S(x)$ in (\ref{bc_scalar}) fixes the rescaling (\ref{rescaling}). 
The constants $b, B_2$ are a priori independent. They encode  the deviation with respect to the vacuum 
solution (\ref{analytic}) where the relation between them is fixed to give  
\be
        B_2 = \frac{2(b x_h-1)^2}{b x_h^3} \ ,
\ee
as found by  eliminating the parameter $a$ from $B'(r_h)$ and $B''(\infty)$ of (\ref{BehaviorExplBHSol}). 
This demonstrates in particular that, to any positive value of $B_2$ 
(i.e. negative $\Lambda$),
two solutions of the form (\ref{analytic}) are available. One of these solutions presents a doubly degenerate 
horizon at $\tilde x_h = x_h + \sqrt{x_h^2 + 8/ B_2}$, corresponding to $a < 0$ in (\ref{analytic}).  

 Our numerical results show  strong  evidence that the analytic solutions can be deformed for generic values
  of $b,B_2$ or, put differently, are just special cases of a much wider family of vacuum solutions
  of the scalar-tensor conformal theory. 
 These new solutions can be characterized by their expansion around the horizon , 
 \be
        B(x) = b(x-x_h) + \frac{b_2}{2}(x-x_h)^2 + \frac{b_3}{6} (x-x_h)^3 + \dots \ \ , \ \ 
        S(x) = 1 + s(x-x_h) + O((x-x_h)^2)
 \ee 
as well as by their asymptotic behavior
\be
       B(x) = \frac{B_2}{2} x^2 + B_1 x + B_0 + \frac{B_{-}}{x} + O(x^{-2}) \ \ , \ \ S = \frac{S_1}{x} + O(x^{-2})
\label{AsymptB-S}
\ee 
where the parameters $B_2$, $b$ have to be set in the boundary conditions while the parameters $B_1,B_0,B_{-}$
and $S_1$ of the scalar field  can be determined from the numerical solutions.

Due to the decay $S \sim 1/x$, the combination $B S^2(x)$ approaches asymptotically
the constant $B_2 S_1^2/2$. This is encouraging since it yields a non-degenerate point particle Lagrangian 
(see Eq. (\ref{L-PointP})) in the asymptotic region.

For $B_2 = 0$, the asymptotic expansion involves 'log'-terms, in particular
 $S(x) \sim (S_1 + S_2 \log(x))/x + O(1/x^2)$
and the expansion of $B(x)$ is more involved.
Black hole solution approaching a de Sitter space-time asymptotically (i.e. with negative $B_2$ or 
positive $\Lambda$) can also be constructed, 
presenting a cosmological horizon at some radius $x=x_c$ with $x_c > x_h$.

We now discuss the new solutions for $B_2 > 0$.

\subsubsection{Case $\nu = 0$}
\begin{figure}[!t]
\centering
\leavevmode\epsfxsize=8.0cm
\epsfbox{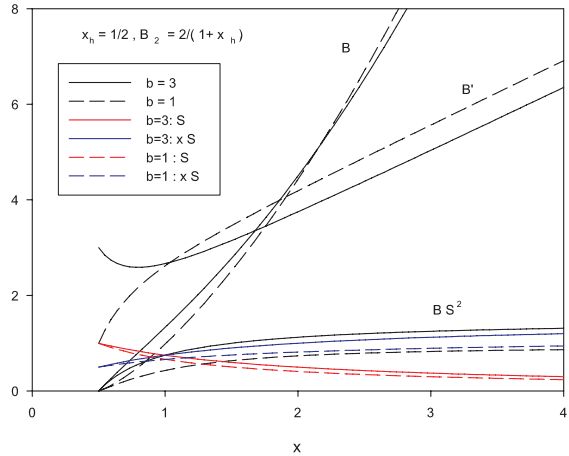}\\
\caption{\label{scalar_2}  \small{First branch vacuum solutions for the case $\nu=0$: Two profiles for 
$B_2=4/3$ with $b=1$ and $b=3$ (analytic solution). Color online 
distinguishes between the curves. In a B\&W version notice that $S(x)$-curves are the two decreasing ones,
while those of $B(x)S^2(x)$ start on the $x$-axis.}}
\end{figure}

We first discuss solutions in the case $\nu = 0$. Two such solutions are presented 
in Fig. \ref{scalar_2} for $x_h=0.5$, $B_2 = 2/(1+x_h)=4/3$.
Here  the analytic solution corresponding to 
$b=3$  is compared to the numerical solution corresponding to $b=1$. 

A natural question consists of determining
the domain of existence of the solutions with a fixed $x_h$ in the plane $b,B_2$.
Our numerical investigations reveal that, for fixed $x_h$ and $B_2$, black holes exist for 
$0 \leq   b \leq  b_{max}$ where the maximal value  $b_{max}$ depend on $x_h, B_2$.
In the limit $b \to 0$, the horizon becomes extremal.
For $x_h=0.5$, we find  $b_{max} \approx 3.7, 3.06, 2.3 $ respectively for $B_2 = 7/3,4/3,0$.
The values of $b$ corresponding to the analytic solutions are
 \be  
b \approx \{1.17 \ , \ 3.41 \} \ \ {\rm for} \ \ B_2 = 7/3 \ \ , \ \ 
b =   \{\frac{4}{3} \ , \ 3 \} \ \ {\rm for} \ \ B_2 = 4/3 \ \ , \ \ 
b = 2 \ \ {\rm for} \ \ B_2 = 0 
 \ee  
The numerical solutions therefore exist for larger values of the parameter $b$ than the analytic ones. 
The following table summarizes these results.
\begin{table}[!h]
\begin{center}
\begin{tabular}{|c|c|c|c|}
\hline \hline 
     &   $B_2 = 7/3$ &  $B_2 = 4/3$ &  $B_2 = 0$ \\
     \hline \hline 
 $b_{max}$               & 3.7  &  3.06 &  2.3 \\
 \hline  $b_{analytic1}$ & 3.41 &  3    &  2 \\
 \hline  $b_{analytic2}$ & 1.17 & 4/3   &  2 \\
\hline
\end{tabular}
\end{center}\vspace{-0.5cm}
\caption{\label{table} \small{Summary of results for the parameter $b$ for three values of $B_2$. }}
\end{table}

\begin{figure}[!b]
\centering
\leavevmode\epsfxsize=8.0cm
\epsfbox{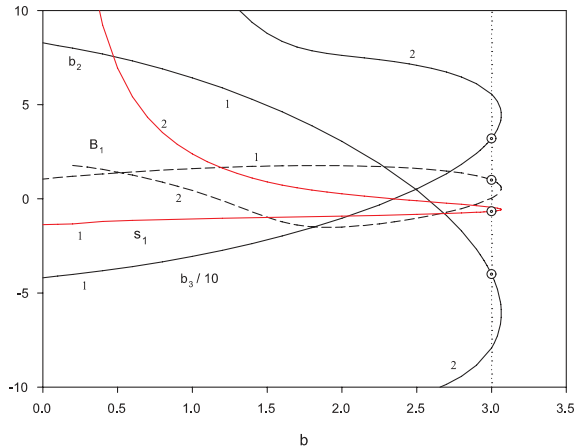}\\ 
\caption{\label{scalar_3}  \small{The two branch structure of the $\nu=0$ solutions. 
 The bullets indicate the corresponding analytic solutions. For the meaning of the various parameters, see text.}}
\end{figure}

\begin{figure}[!t]
\centering
\leavevmode\epsfxsize=10.0cm
 \includegraphics[height=.26\textheight,width=.48\textwidth]{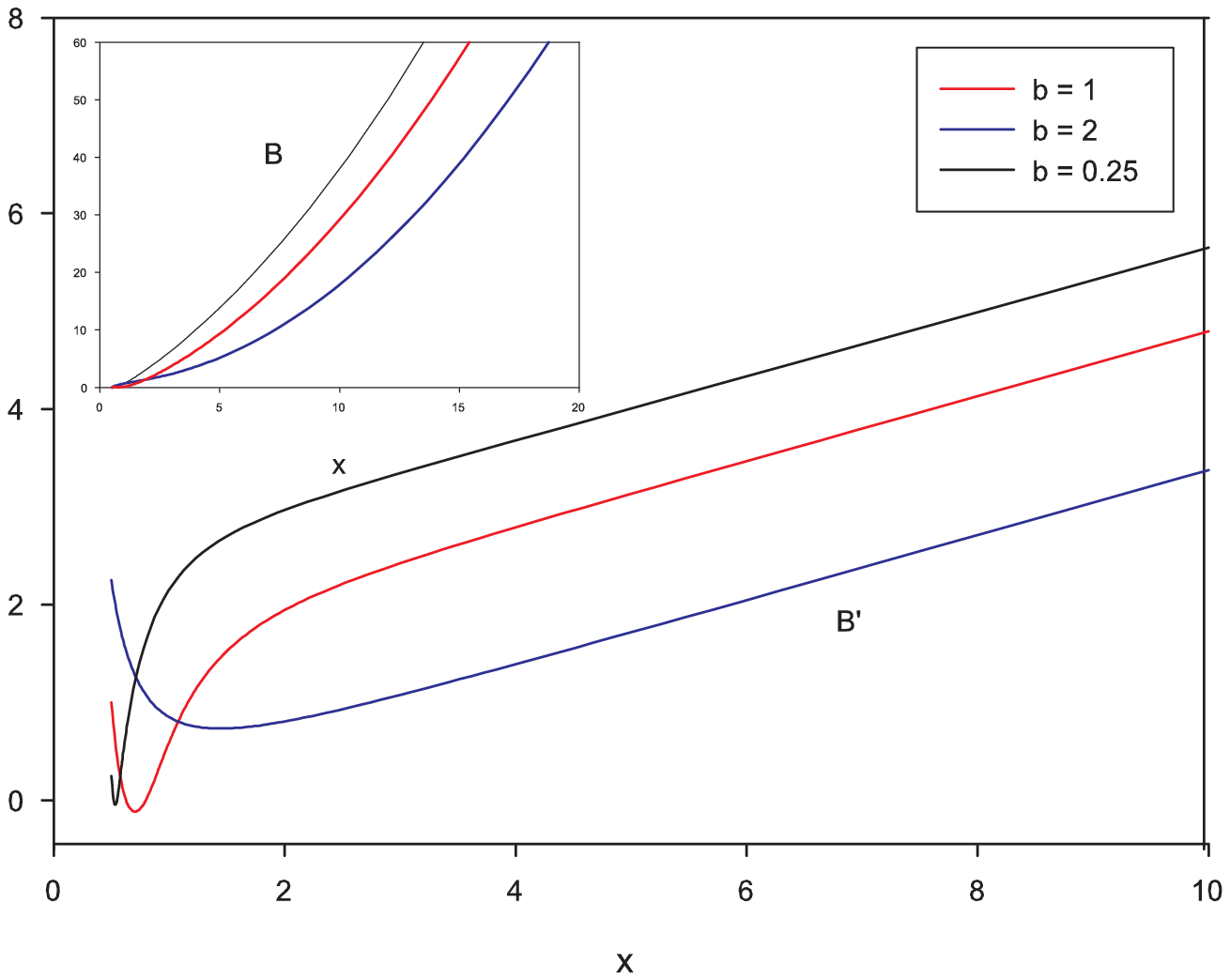}  
 \includegraphics[height=.26\textheight,width=.48\textwidth]{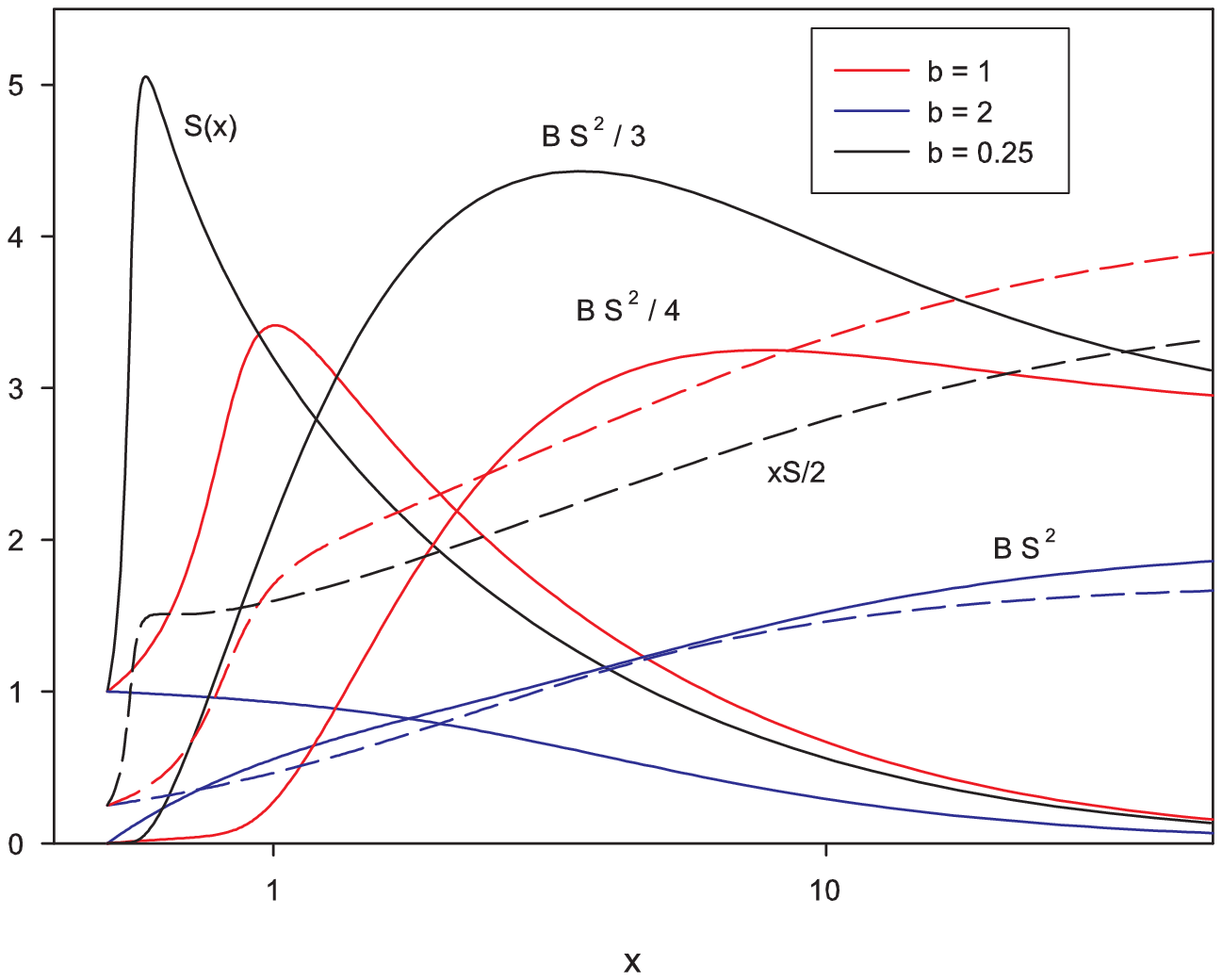}\\
 (a)\hskip 7.5cm (b)\\
\caption{\label{compa_b+s}  \small{Second branch vacuum solutions for the case $\nu=0$:
Three profiles for $B_2=1/3$ with $b=2$, $b=1$ and $b=0.25$. (a) curves for $B(x)$ and $B'(x)$; 
(b) curves for $S(x)$, $xS(x)$ and $S^2(x)B(x)$. }}
\end{figure}

The parameters $b_2,b_3,B_1$ are plotted as functions of $b$  in Fig. \ref{scalar_3}
for $B_2 = 4/3$ (branches labelled `1'). 
The evolution of the parameters $b_2,b_3$ clearly determines the critical phenomenon stopping
the solution at $b=b_{max}$. The property that solutions do not exist for $b > b_{max}$
suggests that a new branch of solutions should occur for $b < b_{max}$, joining the first branch
in the limit $b \to b_{max}$. This was confirmed by the numerics: we indeed managed to construct
a second family of solutions presenting this property. The corresponding data is
presented in Fig. \ref{scalar_3} by the lines labelled with a symbol `2'.  
Decreasing the parameter $b$ along the second branch, we observe very peculiar properties.
     In particular the functions $S(x)$ and $B(r)$ stop to be monotonically decreasing, but present
     respectively a local minimum and a local maximum at two different radii which are rather close
     to the horizon. For $b \to 0$, the position of the local extrema slowly move to the horizon
     and result in large variation of the derivatives $S'(x),B''(x),B'''(x)$ in the region of the
     horizon. The numerical results suggest strongly that the solutions tend to
     a configuration where $S(x)$ presents a singularity at the horizon.
     This appears on Fig. \ref{scalar_3} where the parameters $b_2,b_3, s$ are plotted as functions of $b$.

Profiles of three solutions of the second branch are shown in Fig. \ref{compa_b+s}. Note that the 
effective metric exhibits a similar behavior than above and very different from purely tensorial
CG: the gravitational potential which is encoded in $S^2(r)B(r)$ increases
much less steeply and tends asymptotically to a constant, the spacetime seems to have only a bounded extension 
since the circumferential radial distance $rS(r)$ has a finite limit as $r\rightarrow \infty$ as well as 
the proper radial distance  $\int drS(r)/\sqrt{B(r)}$.

\begin{figure}[!b]
\centering
\leavevmode\epsfxsize=10.0cm
 \includegraphics[height=.28\textheight,width=.54
 \textwidth]{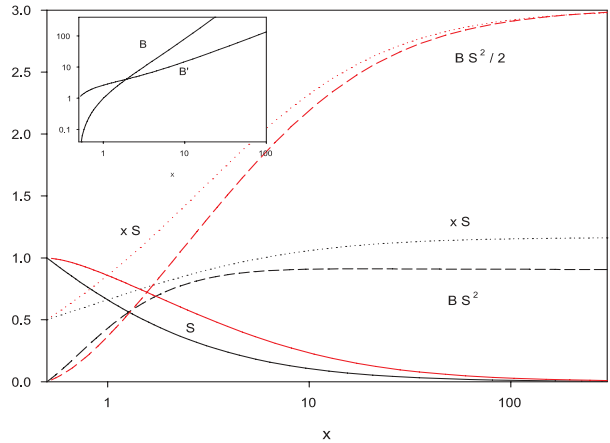}  
\caption{\label{PosNU}  \small{A typical solution with $\nu>0$ (online red) compared 
with a $\nu=0$ (online black) solution. }}
\end{figure} 

\begin{figure}[!t]
\centering
\leavevmode\epsfxsize=10.0cm
 \includegraphics[height=.22\textheight,width=.46\textwidth]{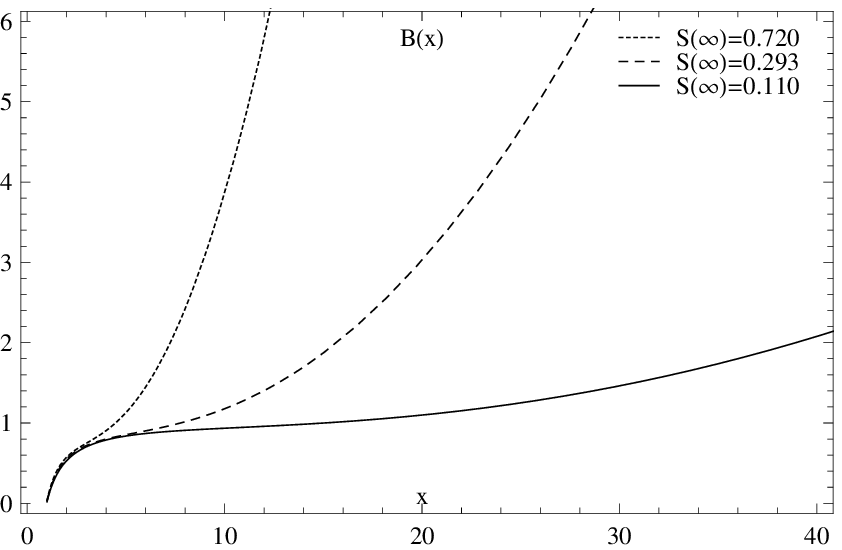}  
 \includegraphics[height=.22\textheight,width=.46\textwidth]{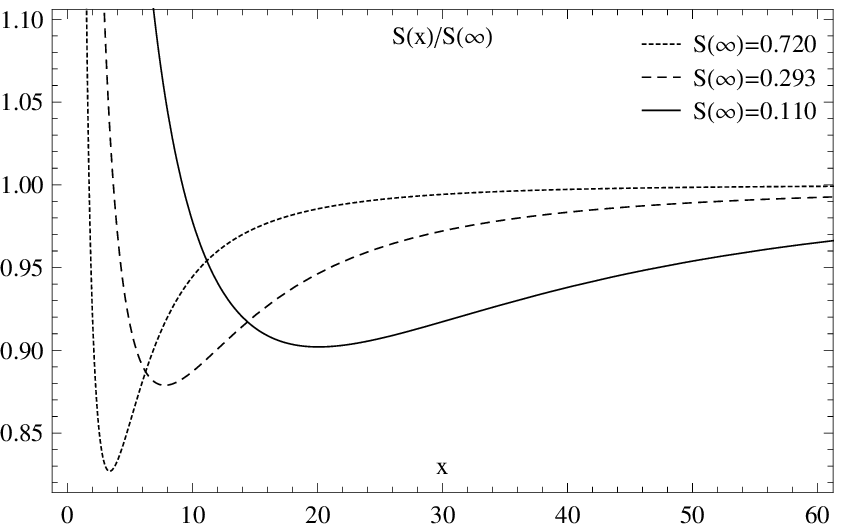}\\
 (a)\hskip 7.5cm (b)\\
\caption{\label{TSOpenSol}  \small{ Open solutions of the vacuum 
scalar-tensor system with $\nu=0.2$ for three asymptotic values of the scalar field.}}
\end{figure} 

\subsubsection{Case $\nu \neq 0$}\label{nuneq0}

For $\nu \neq 0$, the analytic solutions with fixed $x_h,B_2$  are real as long as the condition
\be
  B_2^2 x_h^2 + 6 B_2 \nu x_h^2 + 8B_2 - 3 \nu^2 x_h^2 - 8 \nu \geq 0
\ee
holds. This defines bounds of the parameter $\nu$.
We have also tried to deform the numerical solutions available for $\nu = 0$ to the case $\nu \neq 0$.
The features of the solutions are basically similar. Keeping the parameters $b,B_2$ fixed and
increasing $\nu$, it turns out that the coefficient $S_1$ of the scalar field (defined in 
Eq. (\ref{AsymptB-S})) increases rapidly
and diverges when the coupling constant $\nu$ approaches a critical value.
For example, setting $b=1, B_2 = 4/3$, we find that the main branch exists for $0 \leq \nu \leq 1.65$
while the second branch exists for $0 \leq \nu \leq 0.075$.

Contrary to our expectation, the two solutions available for $\nu = 0$ do not converge to
a common solution while increasing $\nu$ gradually.

All the above mentioned solutions are closed ones. However, there exists another kind of black hole solutions 
which are open and thus of course much more relevant in order to deal with the astrophysical and cosmological issues 
discussed at the introduction. These solutions are characterized (in addition to the horizon) 
by non-vanishing value of $S(\infty)$ 
which fixes the cosmological constant parameter by $\kappa=\nu S^2(\infty)/2$. They have the asymptotic
behavior 
 \begin{equation}
       B(r) \sim  \kappa r^2 + ( B_2 \log (r) +B_1 )r + B_0  \ \ , \ \ 
       S(r) \sim \sqrt{\frac{2\kappa}{\nu}} + S_1 \frac{\log (r)}{r} \ \ , \ \ 
     \label{AsymptB-S-OPEN}  \end{equation}
   where $B_0,B_1,B_2,S_1$ are constants.
Fig. \ref{TSOpenSol} contains a graphic representation of typical solutions. 

\subsection{Regular solutions}
\begin{figure}[!t]
\centering
\leavevmode\epsfxsize=10.0cm
 \includegraphics[height=.26\textheight,width=.48\textwidth]{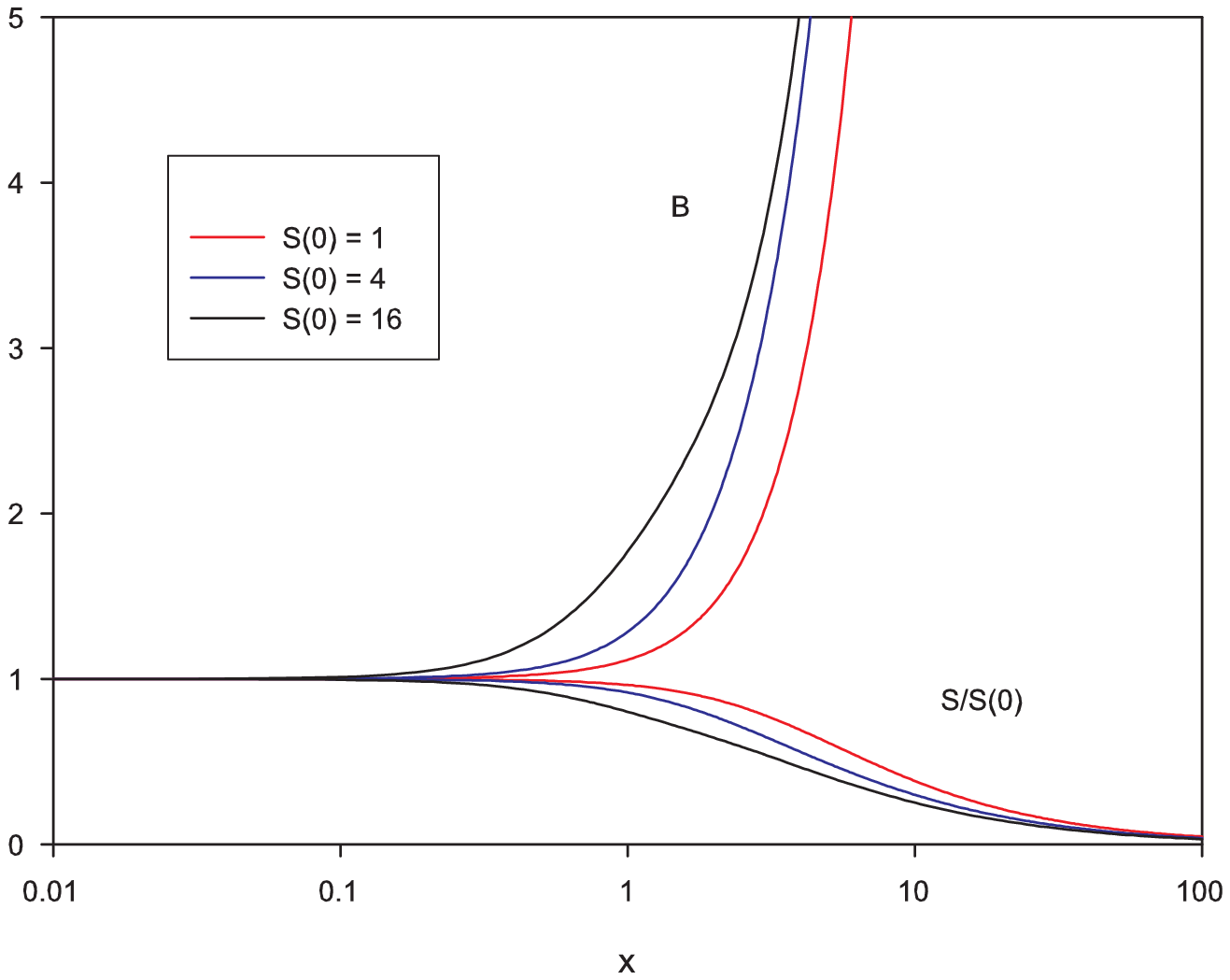}  
 \includegraphics[height=.26\textheight,width=.48\textwidth]{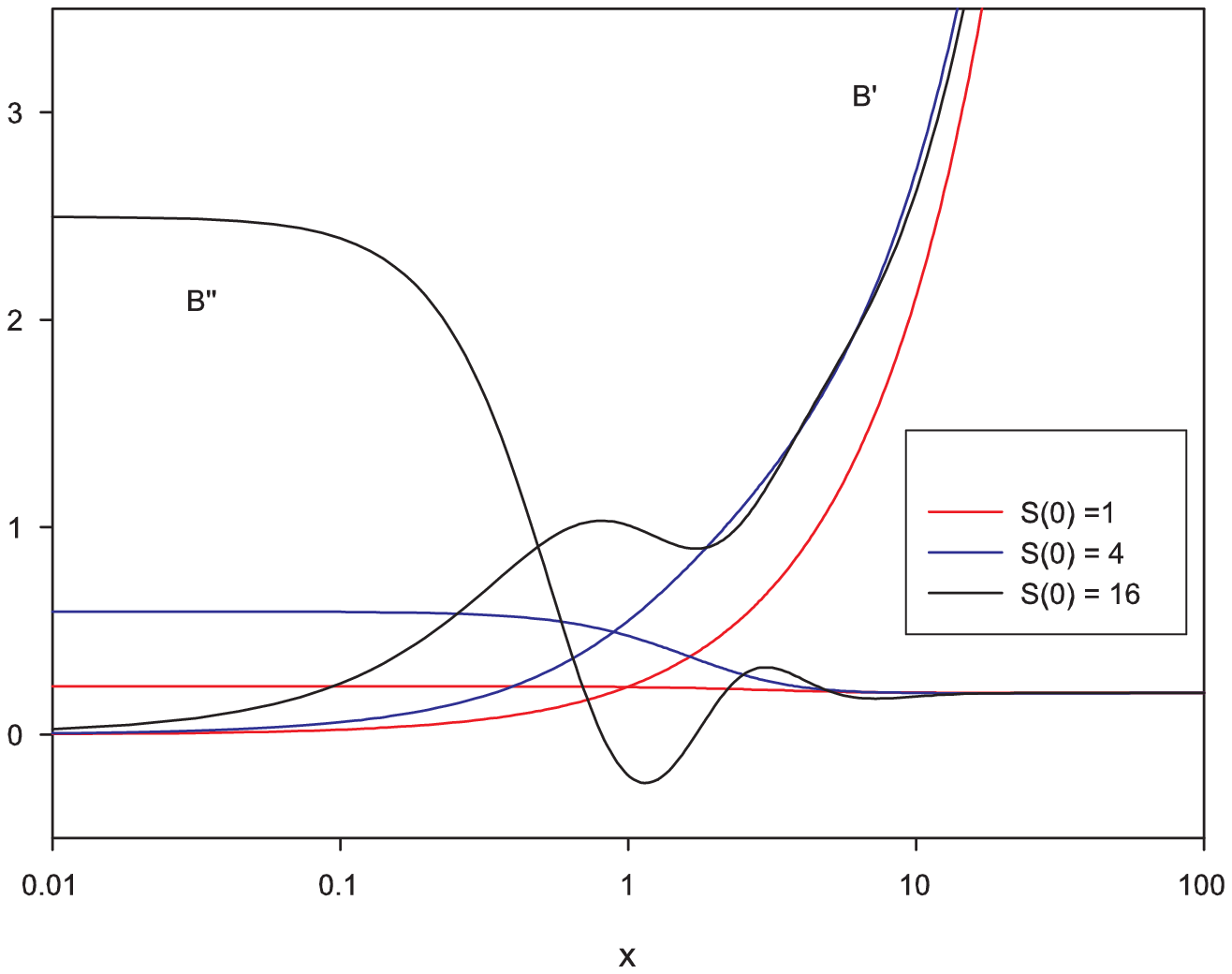}\\
 (a)\hskip 7.5cm (b)\\
\caption{\label{TSRegProf}  \small{Several profiles of the regular solutions of the 
scalar-tensor system with $\nu=0$ for different central value of the scalar field:
$S(0)=1$, $S(0)=4$, $S(0)=16$.  Notice the oscillations for large $S(0)$.}}
\end{figure}

Apart from black hole solutions discussed in the previous section,
the system (\ref{STVacFieldEq}) also admits regular solutions in $r\in [0,\infty]$ which may be viewed
as gravitational solitons in this scalar-tensor theory. Taking advantage of the symmetry (\ref{rescaling}), 
the  boundary conditions can all be fixed to be~:
\begin{equation}
    B(0) = 1 \ \ , \ \ B'(0) = 0 \ \ , \ \ 
  B'''(0) = 0 \ \ , \ \ S'(0)=0 \ \ , \ \ B''(\infty)=2\kappa
\end{equation}
with  $S(0)$ as an additional input which we will allow to vary in a certain range.
It turns out that a family of solutions exists, labelled by $S(0)$. The solutions behave asymptotically 
according to (\ref{AsymptB-S}) and several profiles of such solitons 
are presented in Fig. \ref{TSRegProf}. 

Actually, this kind of solutions must exist also for the boson star system of section \ref{BS}
as completely static self-gravitating solutions -- a novelty in the CG with no analogue 
in standard GR. We have not found them at the time because we used $\omega$ in order to rescale the 
dimensionful  variables. Note however that the interpretation of these static solutions is very different in
both cases: Here it is a gravitational soliton in a scalar-tensor theory, while the same solution within the
purely tensorial CG describes a self-gravitating scalar field, i.e. boson star, 
with the peculiar property that it is purely static. The dotted line in Fig. \ref{TS_BS_1_b}  below shows the 
(inertial) mass of such a static boson star as a function of the central value of the scalar field $S(0)$. 
The oscillatory curve is very similar to the one found for the usual boson stars in
GR \cite{{Jetzer1992},{LeePang1992},{Liddle1992}}.

The striking feature about these regular solutions is that the field $B$ deviates
only a little from the form $B(r) = 1 + \kappa r^2$. The numerical results
indicates that the difference $1-(B''(\infty)/B''(0))$ is positive and of the order of a few percents (we 
checked that this is not a numerical artefact). As a consequence, these 
regular solutions are essentially characterized by their cosmological constant. 

If we examine the family of black holes with a fixed $\Lambda$
and decreasing $r_h$, it turns out that the maximal value $b_{max}$
 of the parameter $b$ increases. 
The numerical results then strongly suggest that the
profile of the regular solution is approached on the interval
$r\in ]0,\infty[$ by the black holes corresponding to the second branch.
The  convergence cannot be extended towards the point at the origin 
because of the different condition of the metric field: $B(r_h)=0$
for black holes, $B(0)=1$ for the soliton. 

\section{Perfect Fluid Solutions in Scalar-Tensor Conformal Gravity} \label{ST-PF}
\setcounter{equation}{0}

 \begin{figure}[!b]
\centering
\leavevmode\epsfxsize=10.0cm
 \includegraphics[height=.26\textheight,width=.48\textwidth]{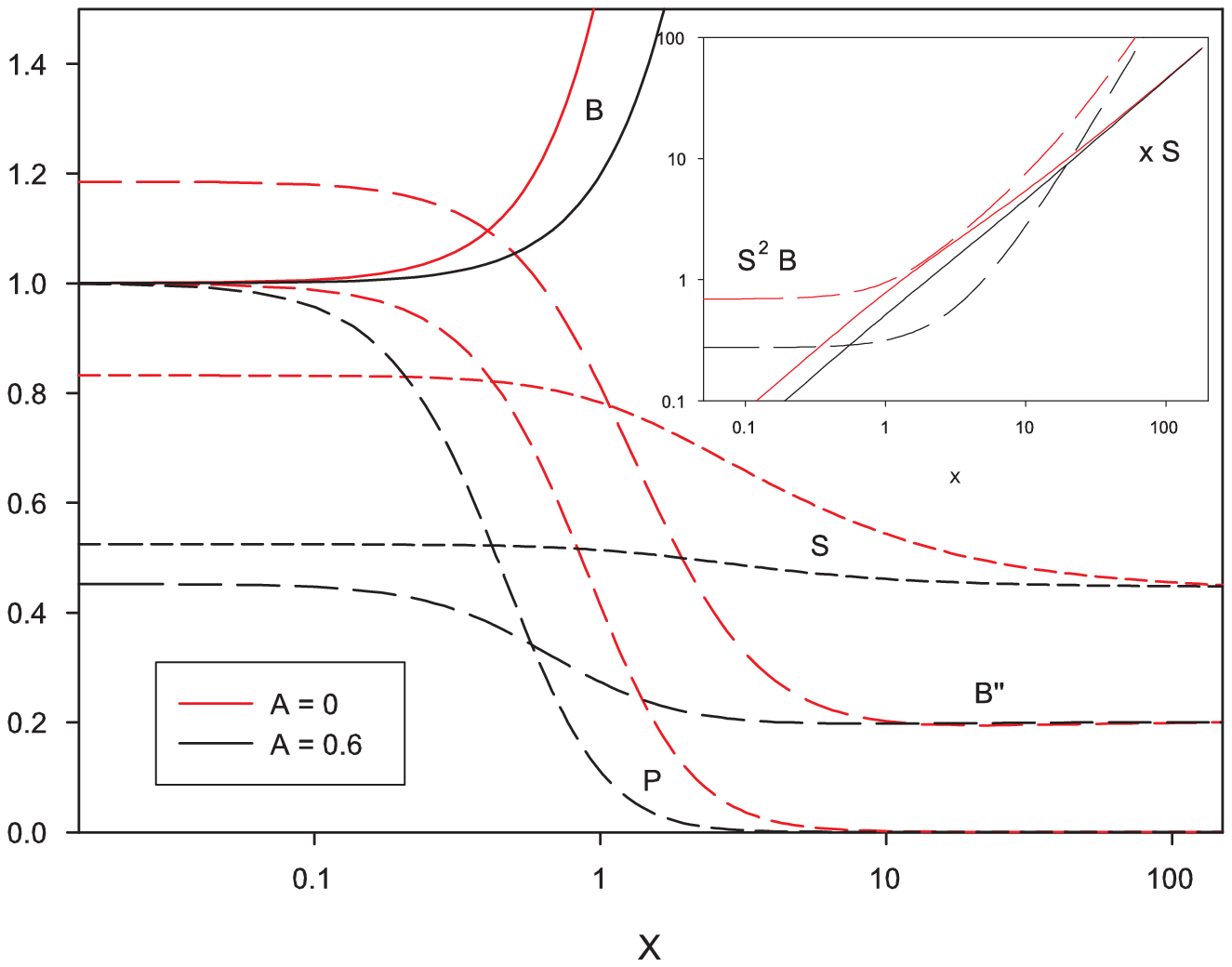}  
 \includegraphics[height=.26\textheight,width=.48\textwidth]{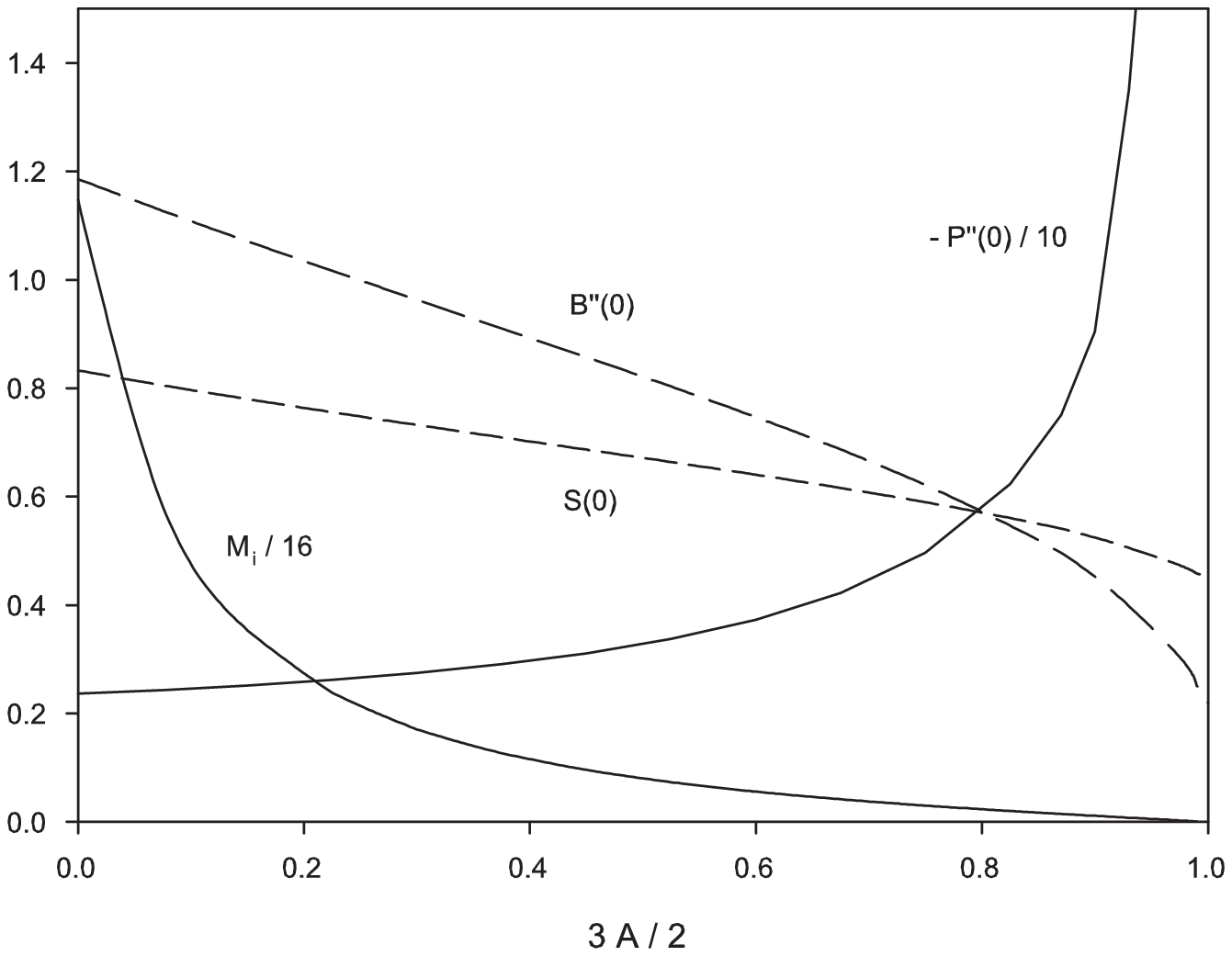}\\
 (a)\hskip 7.5cm (b)\\
\caption{\label{STpoly}  \small{Perfect fluid open solutions in the scalar-tensor theory with 
$\nu=1$. $\kappa=0.1$ :
(a) the profiles of two solutions;  
(b) plots of several characteristics of the solutions as a function of the parameter $A$.}}
\end{figure}  

Having investigated the scalar-tensor vacuum solutions and especially obtaining open solutions,
we now proceed to couple matter sources to this system. The first is the perfect fluid with a 
polytropic equation of state. In this case we are confronted with the following set of equations:
\begin{eqnarray}
\frac{(rB)''''}{r}+\frac{1}{B}\left[ BS'^2 -\frac{\nu}{2}S^4 
+\frac{1}{4}\left(B'+\frac{2B}{r}\right)(S^2)'-\frac{R}{12}S^2
\right]= -\frac{3\alpha}{2B}(\rho + P_r) 
\label{STSphGravFieldEq}
\end{eqnarray}
\begin{eqnarray}
\frac{\left(r^2BS'\right)'}{r^2} - \frac{R}{6} S -\nu S^3 = 0 
\label{STScalarGFieldEq}
\end{eqnarray}
which are supplemented by the conservation law (\ref{ConservEqPF}).

   We have found two types of regular solutions to the equations above distinguished by $S(\infty)$ being
   either zero or non-zero. This corresponds to closed or open spacetime geometries respectively.   
   For the two cases, the field $B(r)$ satisfies the same boundary conditions as in the purely tensorial 
   Conformal Gravity, namely
   \begin{equation}
        B(0)= 1 \ \ , \ \ B'(0)=0 \ \ , \ \ B'''(0)=0 \ \ , \ \ B''(\infty)=2 \kappa
   \end{equation}
   The boundary conditions on the function $S(r)$ are different for the two solutions. Setting $\nu = 0$, we find solutions
   with the asymptotic behavior 
   \begin{equation}
       B(r) \sim \kappa r^2 + B_1 r + B_0  \ \ \ , \ \ \ S(r) \sim \frac{S_1}{r} \ \ , \ \ P_{r}(r) \sim \frac{P_1}{r^4}
   \end{equation}
   The space-time associated with these solution is closed since the function $rS(r)$ varies on a 
   finite range. Similarly, the proper radial distance is bounded from above.
   The solutions of this type can be deformed for $\nu >0$;       
    however, they do not exist for large values of $\nu$. 
   The function $S(r)$ indeed develops a singularity at a finite value of $r$
   when $\nu$ approaches a critical value $\nu=\nu_c$.

   The solutions of the second  type that we constructed are open and exist for generic non-zero values of the coupling 
   constant $\nu$; they are characterized by $S(\infty) > 0$ and  obey asymptotically
   \begin{equation}
       B(r) \sim  \kappa r^2 + ( B_2 \log (r) +B_1 )r + B_0  \ \ , \ \ 
       S(r) \sim \sqrt{\frac{2\kappa}{\nu}} + S_1 \frac{\log (r)}{r} \ \ , \ \ 
       P_r (r) \sim \frac{P_1}{r^4}
   \end{equation}
   where $B_0,B_1,B_2,S_1$ are constants. This form was checked both analytically and numerically.
   The corresponding space-time is open since $r S(r)$ is unbounded from above. Let us point out 
   two features of these solutions
   (i) They do not possess a regular limit for $\nu\to 0$, as seen e.g. from  
   $S(\infty)= \sqrt{2\kappa/\nu} $. 
   (ii)  Non analytical terms ($\log$ terms) appear in the asymptotic expansion
   of the fields $B$ and $S$. These terms seem to be related to the fact that the field $S$ does not go to zero 
   for $r\to \infty$.
    
\begin{figure}[!b]
\centering
\leavevmode\epsfxsize=8.0cm
\epsfbox{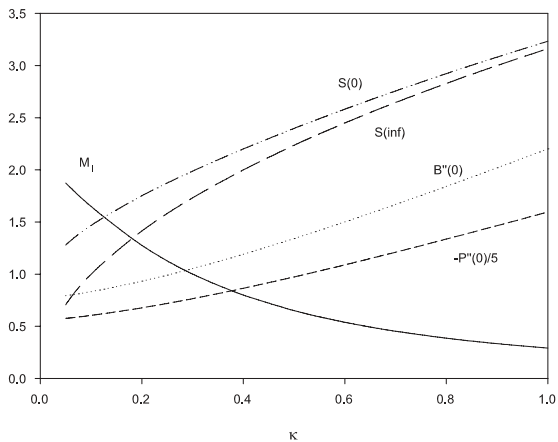}\\
\caption{\label{PolytkappaDep} \small{Perfect fluid solutions in the scalar-tensor theory:
dependence on the parameter $\kappa$. The other parameters are: $\nu = 0.2$, $A = 1/3$. }}
\end{figure}
  

Fig. \ref{STpoly}a shows the
profiles of two solutions with $\nu=1$. The functions $B(r)$ and $P_{r}(r)$ are quite similar to those in the 
``pure tensor''  theory. The main difference is that point particles are now consistently coupled to the gravitational 
field through the new field $S(r)$. The coupling 
is described now by the combination $S^2(r)B(r)$ and it is obvious that besides the cosmological $r^2$ behavior, 
we recover the linear potential with logarithmic modifications. Further study is required in order to check the 
relation of this new kind of solutions to observational data.  Fig. \ref{STpoly}b presents the dependence on the
polytropic index $A$ of the main properties of the solutions. These properties as a 
function of the cosmological constant parameter $\kappa$ are shown in Fig. \ref{PolytkappaDep}.

\section{Boson Stars in Scalar-Tensor Conformal Gravity}
\setcounter{equation}{0}

Next we move to the complex scalar field, i.e. boson stars. 
In this case, the Lagrangian density is the sum of all the previous terms with a possible 
additional coupling between the ``gravitational scalar field'' $S(x)$ and the other scalar $\Phi$,
namely $-\mu S^2 |\Phi|^2/2$ with $\mu$ dimensionless (real) parameter. The field equations will contain
two second order equations for the two scalar fields
\begin{equation}
\frac{\left(r^2Bf'\right)'}{r^2}+
\left(\frac{\omega^2}{B} - \frac{R}{6} - \mu S^2\right)f -\lambda f^3 =0 
\label{SphScFieldEq2}
\end{equation}
\begin{equation}
\frac{\left(r^2BS'\right)'}{r^2} - (\frac{R}{6} + \alpha\mu f^2)S -\nu S^3 = 0 
\label{STSphScalarFieldEq}
\end{equation}
and a fourth order equation for $B(r)$
\begin{eqnarray} \nonumber
\frac{(rB)''''}{r}+\frac{1}{B}\left[ BS'^2 -\frac{\nu}{2}S^4 
+\frac{1}{4}\left(B'+\frac{2B}{r}\right)(S^2)'-\frac{R}{12}S^2
\right] =\\ 
 -\frac{\alpha}{B}\left[ \frac{2\omega^2 f^2}{B}+ Bf'^2 -\frac{\lambda}{2}f^4 - \mu S^2 f^2
+\frac{1}{4}\left(B'+\frac{2B}{r}\right)(f^2)'-\frac{R}{12}f^2
\right] .
\label{STBosonStarFieldEq}
\end{eqnarray}

The space of solutions is quite large  and defined by two types of parameters: those which appear in the 
field equations namely, $\nu$, $\lambda$ and  $\mu$ and parameters 
(integration constants) which specify the solutions like  $S(\infty)$, $\kappa$ etc. 

A systematic survey of all possible solutions is beyond the scope of this work. Here we present the main
properties of several families of solutions in limited but typical regions of parameter space. Here too we 
find closed as well as open solutions. 

We addressed the system (\ref{SphScFieldEq2})-(\ref{STBosonStarFieldEq}) numerically. We 
started by fixing the different coupling constants according to $\nu=\mu=0$, the constant $\alpha$
can then be set to $\alpha=1$ by a rescaling of $f$.  A rescaling of the
radial variable and of the field $S(x)$ allows one to fix $\omega=1$. 
In the reduced system fixed this way, we further assumed $\lambda=1$. 
Regular solutions  to the equations can then be constructed with the following 
boundary conditions~: 
\begin{equation}
B(0)=1 \ , \ B'(0)=0 \ , \ B'''(0) = 0 \ , \ B''(\infty) = 2 \kappa
\end{equation}  
for the metric function, and
\begin{equation}
S(0)=S_0 \ , \ S'(0)=0 \ , \ f'(0) = 0 \ , \ f(r \to \infty) = \frac{f_2}{r^2} 
\end{equation}
for the two scalar functions. Here $S_0$ is an arbitrary constant.
In the numerical analysis, we set $\kappa=0.1$.  
In the limit $S_0 \to 0$, we have $S(x)=0$ and  the boson star solutions of sec. \ref{BSNum} 
are recovered. We have studied how the boson star solution available in CG
is deformed by the additional scalar field and discovered a rather unexpected pattern
which we now discuss.

\begin{figure}[!t]
\centering
\leavevmode\epsfxsize=10.0cm
 \includegraphics[height=.26\textheight,width=.48\textwidth]{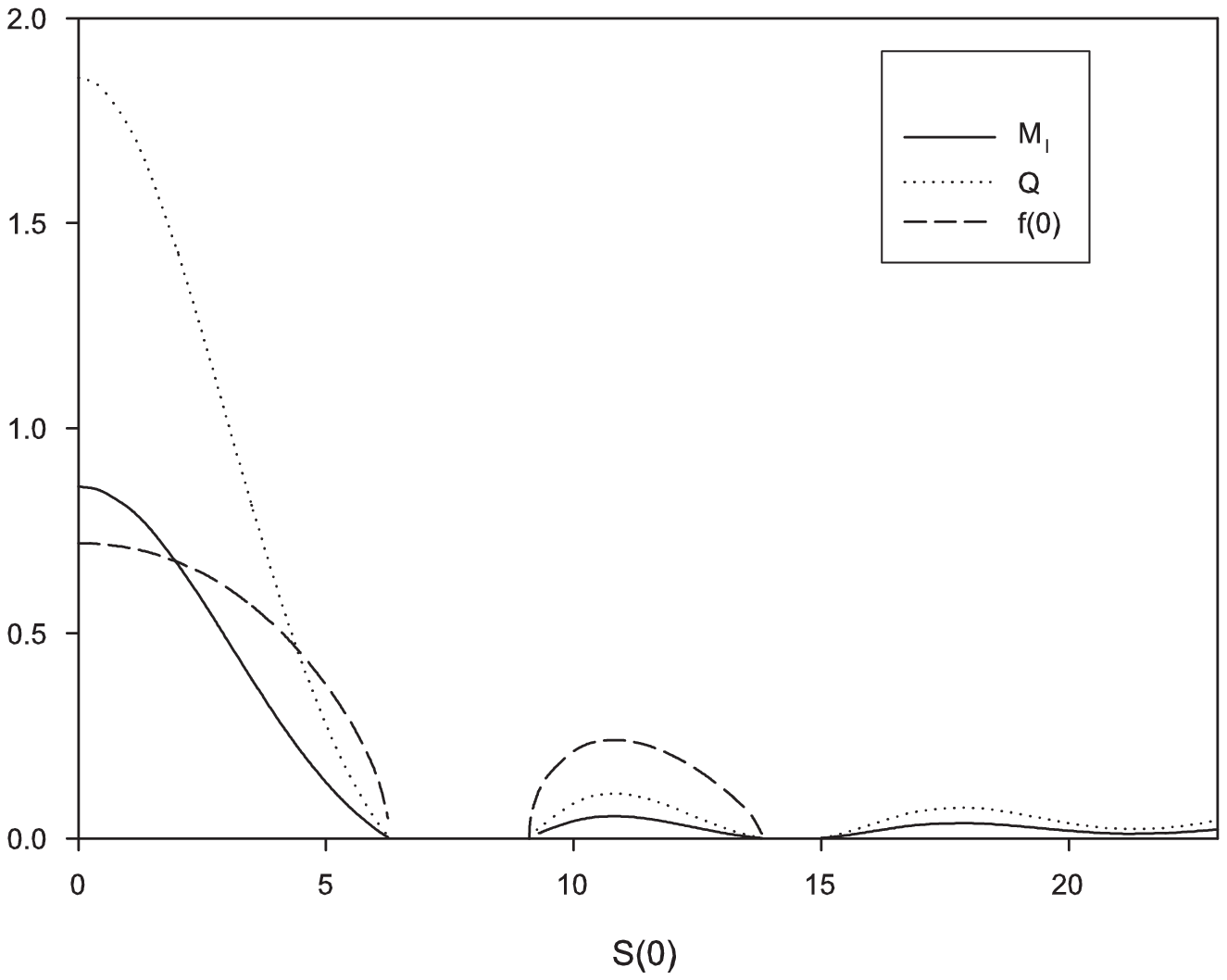}  
 \includegraphics[height=.26\textheight,width=.48\textwidth]{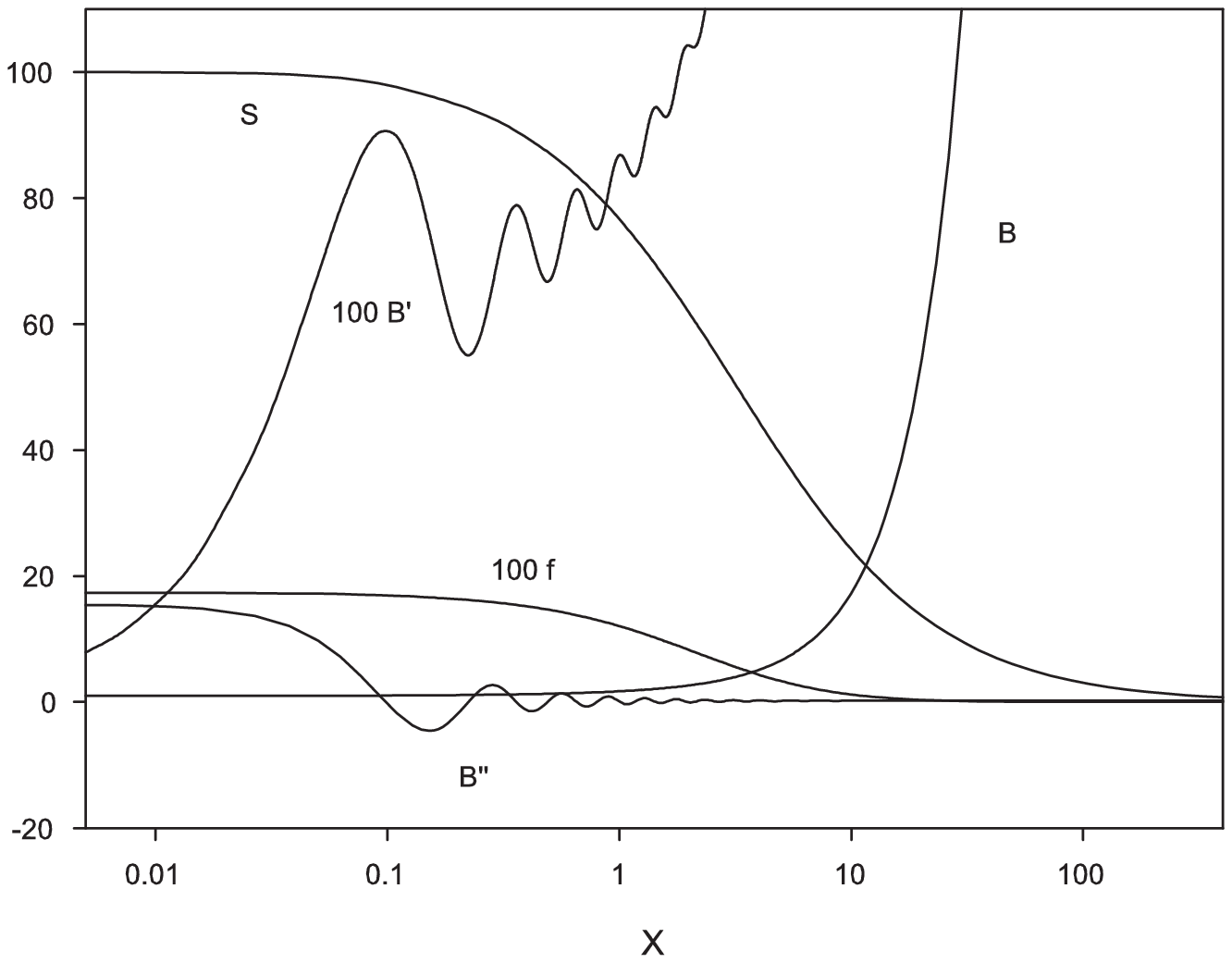}\\
 (a)\hskip 7.5cm (b)\\
\caption{\label{TS_BS_1_a}  \small{Boson stars in the scalar-tensor theory
corresponding to $\lambda = 1$, $\nu=\mu = 0$, $\kappa = 0.1$.
(a) mass, particle number $Q$ and the value $f(0)$ as a function of $S(0)$;  
(b) details of the profile for $S(0)=100$. The high value of $S(0)$
was chosen to get noticeable oscillations - see text.}}
\end{figure}  

\subsection{Closed Solutions}

\begin{figure}[!b]
\centering
\leavevmode\epsfxsize=7.9cm
\epsfbox{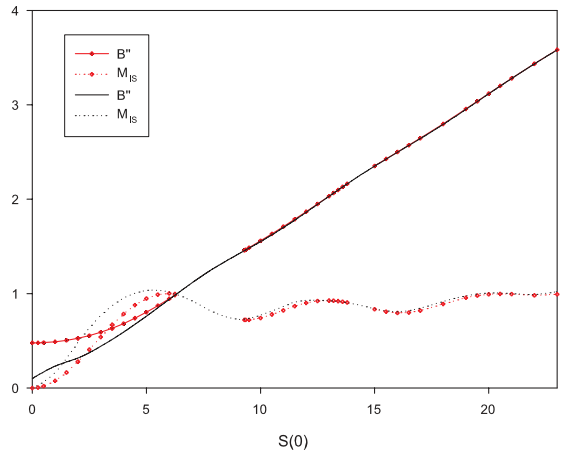}\\
\caption{\label{TS_BS_1_b}  \small{The value $B''(0)$ and the inertial mass of the field S for STBS (on-line red)
and STR soliton (on-line black) as functions of $S(0)$. Notice the gaps in the STBS (red) curves where no 
solutions exist. This can be used to distinguish between the cases in a B\&W plot.}}
\end{figure}

Increasing the parameter $S_0$ gradually, we observe that the boson star gets
continuously deformed by the new scalar field. It turns out that the scalar field of
the boson star tends uniformly to the null function for some critical
value $S_0=S_c$; with our values of the coupling
constants, we find $S_c \approx 6.26$. Accordingly, the mass corresponding to the boson star
(i.e. supported by the field $f(r)$) tends to zero
in this limit, along with the particle number $Q$.
These features are illustrated by Fig. \ref{TS_BS_1_a}a.  
So, for $0 \leq S_0 \leq S_c$ two regular solutions coexist: the scalar-tensor boson star (STBS) and
the scalar-tensor regular solution (STR - the gravitational soliton) with $f(r)=0$ . 
The values of $B''(0)$ and the inertial mass 
 $M_{IS}$  of the  scalar field  $S$  (only) are represented in  Fig. \ref{TS_BS_1_b}.

However this is not the end of the story.
Indeed, while we continue to increase $S_0$, it turns out that STBS solutions 
reappear for $S_0 > 9.1$ and that again, the two non trivial solutions exist
for $9.1 < S_0 < 13.8$. After still another gap of non existence of STBS solutions
they again reappear for $S_0 >15$ and seems then to coexist with the STR (regular) solution
as suggested by Figs. \ref{TS_BS_1_a}a, \ref{TS_BS_1_b}. 

It is challenging to find a full analytical explanation of the features just discussed
above, but so far we have not fully succeeded. Possibly, an explanation is to be found
in the fact that, for large values of $S(0)$ the function $B(r)$ develops some oscillations 
near the origin. These oscillations, which appear more clearly when
looking at $B'(r)$ and $B''(r)$ (see Fig. \ref{TS_BS_1_a}b) are due to a term
\begin{equation}
            Y'' + \frac{S^2}{B}Y + {\rm subdominant \ terms}  = 0 \ \ , \ \ Y \equiv B''
\end{equation}
contained in the equations and which leads to visible oscillations when $S_0$ is sufficiently
large. A detailed numerical study of the ``gap-structured'' phenomenon of the STBS solution shows that:
(i) the function $B''(r)$ is monotonically decreasing for $S_0 < 6.26$; (ii) a local minimum
of $B''$ occur somewhere for $S_0 \in [9.1,13.8]$. This strongly suggests a connection
between the oscillations of the function $B''$ and the pattern of STBS solutions.

In fact, the occurrence of oscillations appears for the STR solution already, i.e. in the absence
of boson star.
This property is illustrated in Fig. \ref{TSRegProf} where the profiles of $B,S,B',B''$
are superposed for three values of $S(0)$.

 \begin{figure}[!b]
\centering
\leavevmode\epsfxsize=10.0cm
 \includegraphics[height=.26\textheight,width=.48\textwidth]{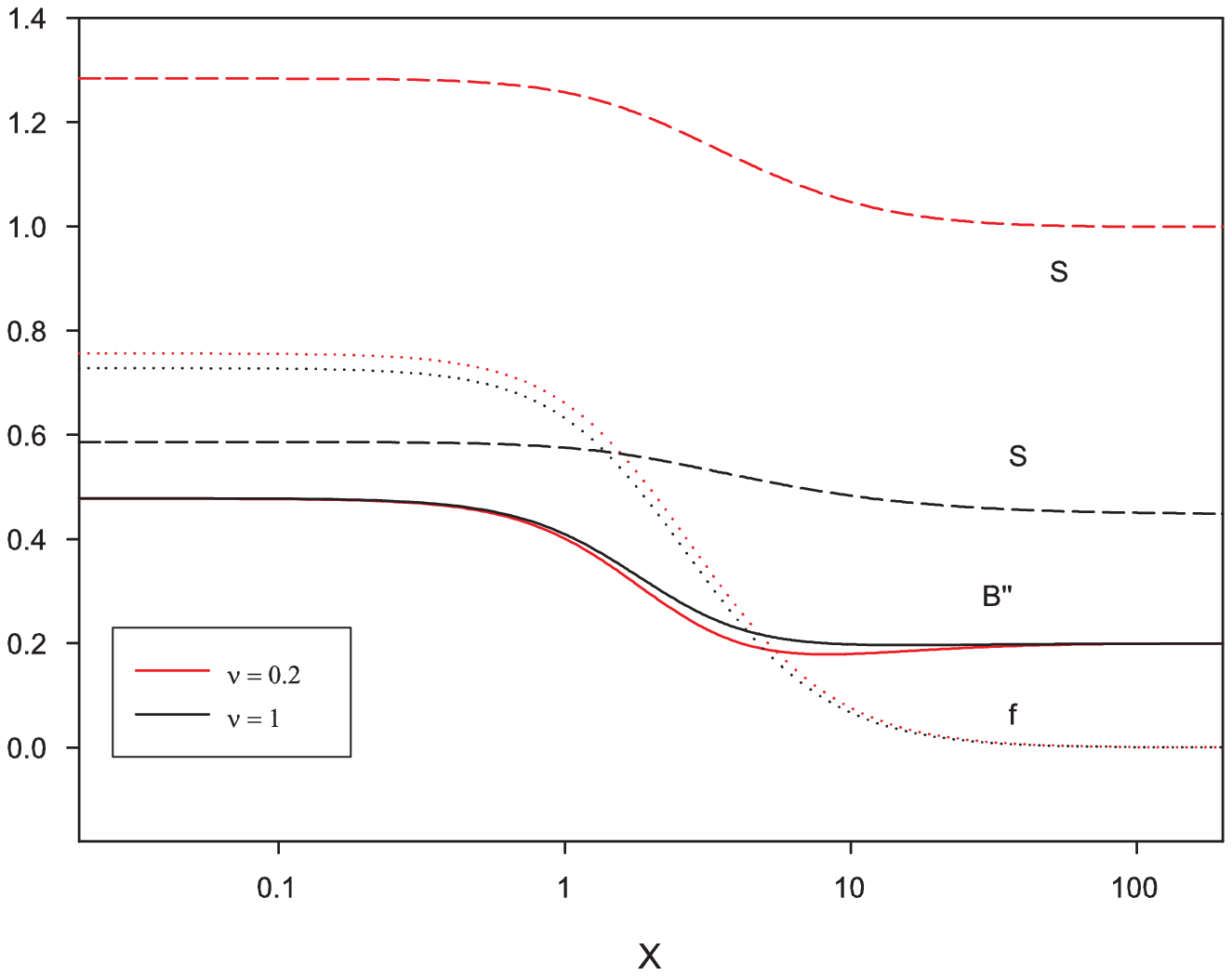}  
 \includegraphics[height=.26\textheight,width=.48\textwidth]{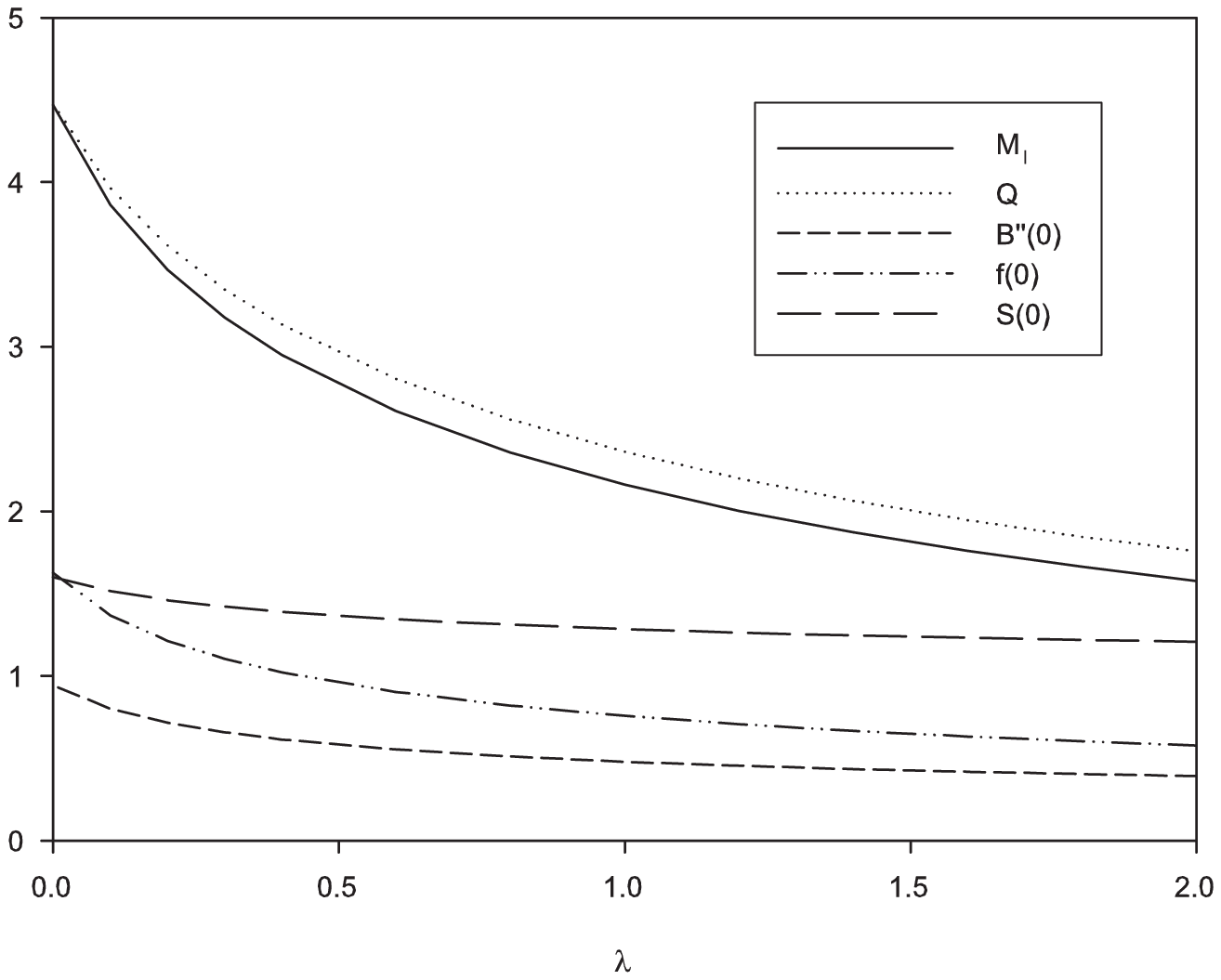}\\
 (a)\hskip 7.5cm (b)\\
\caption{\label{ST-BS-Open}  \small{Open boson star solutions in the scalar-tensor theory:
(a) two profiles for $\nu = 0.2$ and $\nu =1$ with $\lambda = 1$, $\mu = 0$, $\kappa = 0.1$;  
(b) plots of several characteristics of the solutions as a function of the parameter $\lambda$
with $\nu = 0.2$, $\mu =0$, $\kappa = 0.1$. }}
\end{figure}  

\subsection{Open Solutions}
Finally we turn to the very different kind of solutions, namely the open ones. 
Along with the case of polytropes discussed in sec. \ref{ST-PF}, the open solutions are characterized by 
$S(\infty) =\sqrt{2\kappa/\nu}$ and the corresponding asymptotic expansion presents $\log$-terms.
The pattern of solutions of these non linear equations is rich 
   and presents several bifurcations in the space of coupling constants. Fixing 
   the different coupling constants, the solutions are even not
   unique since they are characterized by parameters at the boundary
   (or integration constant like $\kappa$ or $S(0)$), generating
   continuous families of solutions.
   
\begin{figure}[!t]
\centering
\leavevmode\epsfxsize=8.0cm
\epsfbox{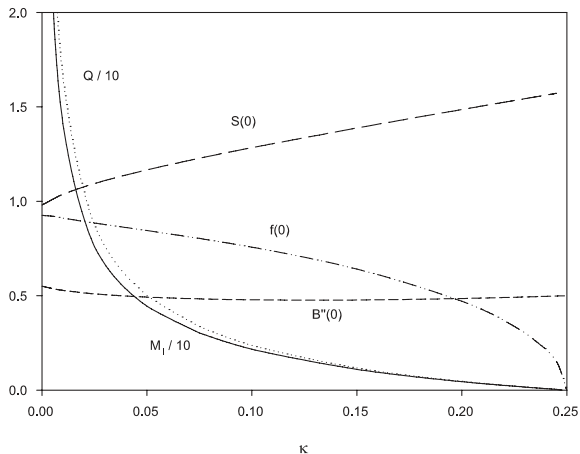}\\
\caption{\label{BSkappaDep} \small{Open boson star solutions in the scalar-tensor theory:
dependence on the parameter $\kappa$. The other parameters are: $\lambda = 1$, $\nu = 0.2$, $\mu =0$.} }
\end{figure}

   Figure \ref{ST-BS-Open}a shows two typical open boson star profiles for two values of $\nu$. 
The quite weak dependence on  $\nu$ makes it sufficient to present in Fig. \ref{ST-BS-Open}b the dependence of the mass, 
charge and other quantities on the self-coupling $\lambda$.  On the other hand, the physical quantities are quite
sensitive to the parameter $\kappa$ as is clearly apparent from Fig. \ref{BSkappaDep}.
  We see that boson stars exist only up to a maximal value of $\kappa$ and that the STBS system bifurcate into a
   regular scalar-tensor solution (STR).
   
\section{Conclusion}
\setcounter{equation}{0}

We have analyzed several types of spherically symmetric solutions in the ``minimal'' CG and
in a scalar-tensor extension. The polytrope solutions in the ``minimal'' case have an 
asymptotically linear gravitational potential and contain a ``wrong sign'' Newtonian component of a $1/r$ 
term only in  asymptotically  anti-de Sitter space which is generated by a ``cosmological integration constant'' ($\kappa$).
These are in line with previous results showing that the generic exterior gravitational fields in this theory have a 
behavior which is very difficult to settle with observations. 

The linear potential may be considered an advantage in the sense that applying 
it in a galactic scale may  provide an explanation for the rotation curves without invoking dark matter.
However, since there is no a-priori reason to refrain from trying CG in a solar system scale, the absence 
of the (attractive) Newtonian potential seems to be a drawback. Of course, it is always possible to claim that 
``conformal polytropes'' made of matter which satisfies $T^\mu_\mu =0$ is a rather special kind of material that
may be found in galactic and intergalactic scale, but this line of argument effectively brings dark matter through the 
``back door''.

From this point of view, a scalar field can be viewed as a more conventional matter source, as scalar fields are 
ubiquitous in theoretical physics. Since the conformal coupling to gravity adds to the energy-momentum tensor terms which 
render the energy density non positive definite, it might be expected that the gravitational fields of boson stars turn out 
to have different behavior. However, we found that boson stars exist only in (asymptotically) anti-de Sitter space, 
otherwise the mass and charge of the  solutions do not converge. Similarly, the gravitational potential of these boson 
stars is linear with a  $+1/r$ additional contribution.

The motivation for the scalar-tensor extension was to search for possible different behaviors. Moreover,
from a formal point of view the scalar-tensor CG is the simplest theory of its kind which 
has physically meaningful time-like geodesics, or in other words, couples consistently to point particles. 

The pattern of the solutions we have found in this case turns out to be very rich  and presents unexpected features,
especially with respect to the purely tensorial CG. First, in the vacuum sector we found Schwarzschild-like 
(or more accurately, Schwarzschild-AdS-like) solutions as well as closed solutions with finite radial extension. 
In addition, there exist regular (``soliton-like'') solutions which have no analogue in ordinary GR.

When matter sources are added, the resulting solutions are classified similarly for both perfect fluid polytropes 
and boson stars: There are closed solutions  which although interesting on their own right cannot 
be considered as relevant in a four-dimensional astrophysical or cosmological context. 
A second type is open solutions with a gravitational potential which 
contains the ``standard'' (by now) linear term modified with logarithmic corrections whose observational relevance 
needs further study. 

The kind of equations we have solved (fourth-order) is  unconventional but could be treated with a good accuracy
by our numerical methods which appear in this case to be indispensable.


\end{document}